\documentclass[lettersize,journal]{IEEEtran}

\usepackage{mathtools}
\usepackage[utf8x]{inputenc}

\usepackage{makecell}

\usepackage{color,soul}
\definecolor{purple}{rgb}{0.5,0,0.5}
\usepackage{subfigure}
\usepackage{moreverb}
\usepackage{epsfig}
\usepackage{amsmath,amssymb,bm,amsthm,mathrsfs,dsfont}
\usepackage{fancyhdr}
\usepackage{adjustbox,lipsum}
\usepackage[font={small}]{caption}
\usepackage{xcolor,soul}
\usepackage{graphicx}
\usepackage{amsfonts}
\usepackage{cite}
\usepackage{microtype}
\usepackage[colorlinks,bookmarksopen,bookmarksnumbered,citecolor=blue,urlcolor=red]{hyperref}
\usepackage{cleveref}

\usepackage{pgfplots}
\pgfplotsset{compat=1.18}
\usepackage{xcolor}
\definecolor{darkGreen}{rgb}{0.47,0.67,0.19}
\definecolor{saddleBrown}{rgb}{0.55,0.27,0.07}
\definecolor{pinkDark}{rgb}{0.72,0.27,1.00}

\usepackage{algorithmicx}
\usepackage{algpseudocode}
\usepackage[norelsize, linesnumbered, ruled, vlined, lined, boxed, commentsnumbered]{algorithm2e}

\newtheorem{remark}{Remark}

\newcommand{\red}[1]{{\color{red} #1}}
\newcommand{\blue}[1]{{\color{blue} #1}}

\allowdisplaybreaks
\usepackage[nodisplayskipstretch]{setspace} 
\begin{document}
\title{ 
Freshness-Aware Constrained Sensing-Aided Beam Prediction with Knowledge Distillation
}
\author{
Abolfazl~Zakeri,~\textit{Member, IEEE}, Nhan~Thanh~Nguyen,~\textit{Senior Member, IEEE},
 Ahmed Alkhateeb,~\textit{Senior Member, IEEE},
 and Markku~Juntti,~\textit{Fellow, IEEE}
 \thanks{A. Zakeri, N. T. Nguyen, and M. Juntti are with the Centre for Wireless
Communications (CWC), University of Oulu, Oulu 90014, Finland, Emails:
\{abolfazl.zakeri; nhan.nguyen; markku.juntti\}@oulu.fi.
\\
A.~Alkhateeb is with the School of Electrical, Computer, and Energy Engineering, Arizona State University, Email: alkhateeb@asu.edu.
\\ This work was supported by the Research Council of Finland through 6G Flagship Program (no. 369116) and projects  DIRECTION (no. 354901) and DYNAMICS (no. 367702).
\\ \indent
The paper's preliminary results were presented at ICASSP 2026 \cite{zakeri_drl_sen} and IEEE ICC 2026 \cite{zakeri_icc26}.
 }
 }
 
\maketitle
\begin{abstract}
Beam prediction leveraging environmental data reduces over-the-air beam training overhead.
Existing frameworks, however, assume continuous access to fresh sensory data, an assumption that breaks down under sensing budget constraints or sensor failures. To make this more practical, this paper proposes a sensing-aided beam prediction framework that operates under an average sensing rate constraint. We incorporate the age of information (AoI) directly into the beam prediction pipeline as a synthetic input modality: the age of the most recently captured data is encoded and fused with visual features through a gating mechanism. This provides the predictor with explicit context information about input sensory data reliability. 
We formalize and examine three fixed sampling policies, accumulated, uniform, and randomized, under the average sensing budget.
We further develop a knowledge distillation (KD) framework that 
operates as a robustness regularizer rather than a pure model compression method. In particular, the high-capacity teacher is trained \textit{unconstrained} on fully sampled data, and its representational knowledge is transferred to a compact student deployed under the sensing budget.
We conduct numerical experiments on the DeepSense 6G data set. The results show that AoI fusion nearly doubles top-1 accuracy at strict sensing budgets, and age-aware models achieve near-optimal top-3 accuracy with only 20\% of the data. Furthermore, we find that the teacher's training regime is a more consequential design choice than the distillation loss~function.
\end{abstract}

\section{Introduction}\label{sec_intro} Millimeter-wave (mmWave) communications offers multi-gigabit data rates and has potential for accurate localization at the cost of severe path loss and high sensitivity to blockage. Harnessing these frequencies requires large antenna arrays and highly directional beams, which in turn demand
accurate and low-latency beam management~\cite{Gerhad_bt, multimodal_wc_mag}.
Traditional over-the-air beam management relies on exhaustive codebook sweeping,
which consumes a significant fraction of the available radio resources
as pilot and training overhead. It can also become impractical
in high-mobility scenarios~\cite{R_Heath_mag, Ahmed_vision}. Reducing this overhead
while maintaining reliable beam alignment is a central challenge in the design of mmWave~systems.

A compelling class of solutions substantially reduces over-the-air beam management overhead by leveraging environmental context extracted
from sensors mounted at the base station~(BS) or user
equipment~(UE), broadly referred to as ``multimodal sensing-aided communications". Cameras, global navigation satellite systems, e.g., global positioning system (GPS) receivers,
light detection and ranging (LiDAR) scanners, and radar units can
each provide a partial view of the propagation environment, and
machine-learning (ML) models trained on such \emph{sensing-aided} data have demonstrated accurate beam prediction without exhaustive
channel measurements~\cite{R_Heath_mag, Gerhad_bt, multimodal_wc_mag}. 
Among the various sensing modalities, vision-based approaches using red-green-blue
(RGB) images have proven particularly effective, with convolutional neural network (CNN) encoders extracting discriminative features
for beam classification and tracking~\cite{ahmed_vis_tvt, Ahmed_vision}.

Despite the progress, a critical but often overlooked aspect of these sensing-aided beam prediction approaches is the \textit{availability of fresh} sensory data at inference in practice. 
Most existing works implicitly assume that fresh, maximally relevant sensory data is perfectly available on demand at every scheduling slot.
 In practice, however, continuous sensing is limited by power budgets, computational processing capacity, and
communications overhead for transmitting or storing sensory data, as well as the possibility of sensor failures or intermittent
connectivity. A vehicular user traversing a coverage area may
encounter periods during which the camera feed is temporarily unavailable, or the network's sensing budget is exhausted~\cite{Tan_V2X_25}.
Under such conditions, the beam prediction must rely on data sampled at
some earlier times, that is, \emph{stale} observations, whose relevance to the current beam degrades with the elapsed time. This
\emph{data freshness} dimension is absent from the vast majority of
existing sensing-aided beam prediction frameworks, rendering their
performance assessments overly optimistic for realistic~deployments.

To address the gap, we build on the \emph{age of information}~(AoI)
concept~\cite{Roy_2012, AoI_Monograph_Modiano}, which defines the time elapsed since the most recent observation was acquired, as a principled metric to quantify and manage data staleness. We formulate a constrained
multimodal sensing-aided beam prediction problem in which a (normalized) sensing
budget $\alpha^{\max} \in (0, 1]$ limits the fraction of time slots in which
fresh data may be obtained. Rather than treating AoI solely as a
scheduling objective, we incorporate it directly into the deep neural-network-based beam prediction
pipeline as a \emph{synthetic modality}: the age of each available
image sample is computed at inference time, encoded, and fused with
the image features through a gating mechanism~\cite{WenTan2025_gatefus} before
being passed to the beam predictor. This transforms the AoI from a
passive constraint descriptor into an active, discriminative feature
that informs the predictor about the reliability of its visual
inputs. A gated recurrent unit (GRU) is further used to aggregate
temporal histories of sampled observations, capturing the sequential
structure of the sensing data sequence. Together, these design
choices enable the model to reason explicitly about both \emph{what} it has seen and \emph{how stale} that sensory data is.

We propose to employ knowledge distillation (KD)~\cite{Hinton_KD} to
enable a high-performance compact model under the sensing constraint,
with a fundamentally different motivation than its conventional use
for model compression. Specifically, we exploit the fact that the
teacher model is not limited by the inference-time sensing budget
during training. Hence, it can be trained under unconstrained access
to fully sampled, temporally dense observations, allowing it to
develop rich inter-class beam representations that are simply not
accessible to a model trained on sparse, stale data inputs. These
abstract beam-class relationships are then transferred to the compact
student model via soft-label supervision. This enables the student to
generalize under the same sensing budget at inference far better than
training under the constraint alone would allow. In this setting,
KD's role shifts from compression to that of an implicit regularizer
against the distribution shift induced by the sensing constraint.

Our main contributions are summarized as follows:
\begin{itemize}
    \item \textbf{Constrained multimodal sensing formulation:}
    We pose the sensing-aided beam prediction problem under an average
    sensing rate constraint. This formulation captures the realistic
    trade-off between sensing cost, i.e., data acquisition and
    processing overhead, and beam prediction quality, and can naturally
    be extended to per-modality sensing costs to enable modality
    selection in multimodal sensing-aided communications.

    \item \textbf{Age-aware multimodal fusion:} We propose
    incorporating the AoI of the most recently captured image
    sample as a synthetic modality, fused with CNN image features
    through a gating mechanism~\cite{WenTan2025_gatefus}. We demonstrate that this low-cost, zero-overhead signal provides discriminative
    context about input reliability, yielding significant accuracy improvements particularly under strict sensing~budgets.

    \item \textbf{Knowledge distillation:} We extend the KD
    framework~\cite{Hinton_KD} to the constrained sensing setting,
    where KD here serves not as a model compression method but as a
    bridge between unconstrained teacher training and
    \textit{budget-constrained} student inference. We systematically
    study the impact of distillation loss choice and the teacher
    training regime (unconstrained versus budget-constrained) on
    student accuracy. We observe that the teacher's training regime is a
    more consequential design choice than the specific distillation
    loss function.

     \item \textbf{Fixed sampling policy comparison:} We formulate
    three budget-satisfying sampling policies, accumulated, uniform,
    and randomized, and characterize their frame-selection behavior as
    a function of $\alpha^{\max}$~\cite{Yin_Sun_2}.

    \item \textbf{Comprehensive empirical evaluation:} Experiments
    on the real-world DeepSense 6G dataset~\cite{Ahmed_deepsense} show that AoI fusion alone can nearly
    double top-1 beam prediction accuracy at $\alpha^{\max} = 0.1$
    relative to an image-only baseline, and that age-aware KD
    achieves the best overall accuracy-complexity trade-off.
\end{itemize}

\smallskip
\noindent\textit{Paper organization.}
The remainder of the paper is organized as follows.
Related work is reviewed in Section~\ref{sec_rw}.
The system model and problem formulation are presented in
Section~\ref{sec_sm}.
Section~\ref{sec_method} describes the proposed age-aware ML framework.
The knowledge distillation framework and training procedures are respectively detailed in Sections~\ref{sec_kd} and \ref{sec_training}.
The experimental setup and simulation results are presented in Section~\ref{sec_nr}.
Section~\ref{sec_con} concludes the paper.

\section{Related Work}\label{sec_rw}
\subsection{Sensing-Aided Beam Prediction and Beam Management}
The use of non-radio frequency (non-RF) sensing data to assist beam prediction has attracted substantial research attention over the past several years. Early works established the feasibility of position-aided beam selection~\cite{R_Heath_mag}, demonstrating that even coarse GPS coordinates can reduce the beam codebook search space and decrease training overhead. Vision-aided approaches subsequently showed that RGB cameras, being both low-cost and information-rich, are particularly effective: CNNs trained directly on image data achieve high top-$k$ accuracy on real-world mmWave datasets~\cite{ahmed_vis_tvt, Ahmed_vision}. Multimodal fusion, combining vision with GPS, LiDAR, or radar, has been shown to further improve robustness, especially in scenarios where a single modality may be degraded~\cite{Ahmed_vision, RHeath_BT_multiuser, vision_aid_pos_JSAC_24}. Prototype demonstrations of camera-assisted beam tracking~\cite{multmodal_exp_vtc} and vision-assisted digital twins~\cite{digital_twin} have validated the practicality of vision-aided management in laboratory and simulation settings, while Vuckovic {\it et al.}~\cite{multimo_revis} revisited the evaluation metrics used for multimodal beam prediction to provide a more consistent performance baseline.

Architectural advances have further pushed the frontier of multimodal sensing-aided beam management. Mollah {\it et al.}~\cite{Mollah_multimodal_attention} propose a transformer-based multimodal fusion framework that employs multi-head cross-modal attention to learn inter-modal dependencies for vehicular mmWave beamforming across vehicle-to-infrastructure (V2I) and vehicle-to-vehicle (V2V) scenarios, achieving over 96\% top-15 beam accuracy. Xie {\it et al.}~\cite{Xie_contrast_bt} address the modality alignment problem through contrastive learning, improving feature consistency between heterogeneous sensing streams and yielding higher beam tracking accuracy. Zheng {\it et al.}~\cite{Zheng_JEPA_MSAC} adopt a self-supervised joint-embedding predictive architecture that pretrains a shared backbone on temporal sequences of multimodal measurements, then fine-tunes lightweight heads for beam prediction, localization, and received signal strength indicator (RSSI) estimation simultaneously. For orthogonal frequency-division multiplexing systems under non-line-of-sight conditions, Li and Yu~\cite{Li_OFDM_mm} integrate camera images, LiDAR point clouds, and RF pilots through a graph neural network to optimize beamforming vectors. Active sensing approaches, where the BS adaptively selects probing beams based on accumulated observations and environmental maps~\cite{Cai_mapISAC}, offer a complementary direction to passive multimodal fusion: the sensing policy itself is optimized jointly with the communication objective. Comprehensive surveys have cataloged the rapid growth of this field~\cite{Tan_V2X_25, Gerhad_bt}.

A critical assumption shared by almost all of the above works is that sensing data is always available and fresh at inference time. Sensing constraints, budget limitations, and the temporal variability of data freshness are not modeled. Our work explicitly addresses this gap by formulating the problem under an average sensing rate constraint and incorporating AoI as a dynamic, learnable feature into the beam prediction~pipeline.

\subsection{Knowledge Distillation for Beam Prediction}
Model complexity is a practical barrier to deploying sensing-aided beam predictors at the network edge. KD~\cite{Hinton_KD} offers a principled approach to constructing compact, deployable models: a lightweight student network is trained to match the soft output distributions of a larger, more capable teacher, transferring generalizable knowledge while dramatically reducing parameter count and inference cost. In the wireless communications domain, Ma {\it et al.}~\cite{Ma_KD} propose a CNN+GRU teacher trained on joint camera and radar data from the DeepSense 6G dataset and compress it into a student model that reduces parameter count by over 27 times while retaining 96\% top-5 beam prediction accuracy. Park {\it et al.}~\cite{Walid_ML25} develop a cross-modal relational knowledge distillation algorithm that transfers knowledge from a multimodal (LiDAR+radar+RGB+GPS) teacher to a radar-only student, enabling high accuracy with only 10\% of the teacher's parameters, via a realistic simulation framework built on the CARLA autonomous-driving simulator. 
We have addressed the case where the original training data is unavailable at distillation time, proposing a data-free KD approach for LiDAR-aided beam tracking \cite{zakeri_DF_KD}.

Prior KD work for beam prediction did not consider the interaction between sensing constraints, data freshness, and distillation quality. In particular, the question of how the teacher's training regime, whether under unconstrained or budget-constrained sensing, affects the quality of the soft targets provided to a student deployed under sparse sensing conditions has not been studied. Our paper fills this gap and reveals that the teacher's training regime is a more consequential design choice than the specific distillation~loss~function.
\subsection{Age of Information and Constrained Multimodal Sensing}
The AoI metric was introduced by Kaul {\it et al.}~\cite{Roy_2012} and subsequently analyzed in depth in the context of status update systems~\cite{AoI_Monograph_Modiano, Yin_Sun_2}. Optimal sampling and scheduling policies for minimizing AoI have been extensively studied under diverse network models, including multi-source systems~\cite{Yin_Sun_2} and energy-harvesting networks~\cite{AoI_Monograph_Modiano}. More recently, Zhang {\it et al.}~\cite{Sun_multimodal_RI} studied the joint scheduling problem for multimodal \emph{remote inference}, showing that AoI-aware policies can meaningfully reduce the inference error rate of multi-sensor classifiers, a result that motivates AoI-awareness at the \emph{learning} level, not only the scheduling level. Despite this body of work, the integration of AoI into the \emph{training} of data-driven beam prediction models, rather than solely as a scheduling objective, has not been fully explored.

AoI has not previously been used as a synthetic input feature for neural-network-based beam predictors. The existing scheduling literature optimizes AoI as a \emph{system-level objective}, whereas we treat it as a per-sample \emph{learning signal} that informs the beam predictor about the reliability of its visual context. This distinction is fundamental: our model does not minimize AoI, it learns to \emph{cope with high AoI}, due to, e.g., packet losses or sensor failure, by explicitly conditioning its predictions on input data freshness.
\subsection{Distinction from Prior Work}
The most closely related works to this paper are~\cite{Ahmed_vision}
and our earlier conference papers~\cite{zakeri_drl_sen}
and~\cite{zakeri_icc26}. In~\cite{Ahmed_vision}, Charan \emph{et al.}
developed a multimodal (vision+GPS) machine-learning framework for
mmWave beam prediction in real vehicular scenarios, assuming fully
available fresh data and no sensing constraints.
In~\cite{zakeri_drl_sen}, we addressed the sensing scheduling problem
using deep reinforcement learning, where AoI entered only as a state
and reward component. In~\cite{zakeri_icc26}, we further incorporated
AoI into predictor training through dataset augmentation, replicating
each sample across all age values up to a predefined limit and
concatenating the scalar age with the input features, while retaining
the architecture of~\cite{Ahmed_vision}.

The present work differs from~\cite{zakeri_drl_sen}
and~\cite{zakeri_icc26} in three respects. First, AoI is integrated
architecturally rather than through the training data: a learned age
encoder projects the age of each retrieved sample into an embedding
that modulates the visual features via a dimension-wise gating
mechanism. This removes the predefined age limit, which
in~\cite{zakeri_icc26} must be retuned for each sensing budget,
together with the proportional growth of the training set that the
augmentation entails. The concatenation variant evaluated in
Section~\ref{sec_nr} corresponds to the fusion mechanism
of~\cite{zakeri_icc26}, and is shown to be outperformed by gating
fusion, particularly at low budgets. Second, the predictor operates
on a finite history of \textit{non-consecutive} observations,
determined by the sampling policy and aggregated by a GRU, whereas
\cite{zakeri_drl_sen} and~\cite{zakeri_icc26} use only the most
recent sample and prior sensing-aided works assume a history window
of \emph{consecutive} samples. Third, neither conference paper
addresses model complexity; here, a compact predictor with a
180-fold parameter reduction is trained within an age-aware KD
framework that has no counterpart
in~\cite{zakeri_drl_sen},~\cite{zakeri_icc26}. In addition, whereas
\cite{zakeri_drl_sen} and~\cite{zakeri_icc26} optimize the sensing
schedule for a fixed predictor, this work fixes the schedule to
budget-satisfying policies and instead optimizes the predictor.

The role of KD in this work also differs from its conventional use.
Rather than serving compression alone, KD bridges unconstrained
teacher training and budget-constrained student inference, thereby
acting as a regularizer against the induced distribution shift. This
is distinct from data-free distillation~\cite{zakeri_DF_KD}, where
the training data itself is unavailable. We systematically study the
effect of the distillation loss and the teacher training regime, and
identify the latter as the dominant design factor.
\section{System Model and Problem Formulation}\label{sec_sm}
We consider a downlink mmWave communications system consisting of a BS and a single-antenna mobile UE.
The BS is equipped with $N$ antennas and an RGB camera. The BS adopts
a predefined analog beamforming codebook of $M$ beams, ${\mathcal{F} =
\{\mathbf{f}_1, \ldots, \mathbf{f}_M\}}$, for signal transmission,
where $\mathbf{f}_m \in \mathbb{C}^{N \times 1}$ with
${ \|\mathbf{f}_m\|_2^2 = 1}$, ${m = 1, 2, \ldots, M}$.
\\\indent 
Denote by $\mathbf{h}(t) \in \mathbb{C}^{N \times 1}$ the channel
between the BS and the UE at time slot $t$. Let $x(t) \in \mathbb{C}$
be the transmit data symbol from the BS to the UE at slot $t$, with
$\mathbb{E}\{|x(t)|^2\} = P$, where $P$ is the transmit power.
Suppose the beamforming vector $\mathbf{f}(t) \in \mathcal{F}$ is
chosen at time $t$; the received signal at the UE is then
\begin{equation}
    y(t) = \mathbf{h}^{\mathrm{H}}(t)\mathbf{f}(t)s(t) + n(t),
    \label{eq:received_signal}
\end{equation}
where $n(t) \in \mathbb{C}$ is additive white Gaussian noise distributed as $\mathcal{CN}(0, \sigma^2)$, with $\sigma^2$ denoting the noise variance at the UE. 

In conventional beam scanning, the best
beam $m$ at time $t$ can be chosen to maximize the received
signal-to-noise ratio~(SNR)
    $\dfrac{|\mathbf{h}^{\mathrm{H}}(t)\mathbf{f}_m(t)|^2}{\sigma^2}$.
This typically requires channel state information or exhaustive search-based
beam training, both of which are costly in mmWave systems. 
We therefore consider a data-driven \textit{sensing-aided} beam
prediction approach to reduce the CSI acquisition/beam training overhead.

In this paper, by sensing we specifically refer to the acquisition of RGB
image data, together with its associated preprocessing pipeline. This visual data
acquisition, however, cannot generally be realized at every time slot
in practice. For instance, in a distributed integrated sensing and
communication deployment, multiple sensing nodes must relay
their locally captured observations to a central processing unit over
a shared, capacity-limited backhaul link; even if each node could
capture continuously, only a fraction of these sensory observations can be
transmitted and processed within a given beam decision interval. Analogous
constraints arise even in a single-sensor deployment, from limited
on-device sensing power, computational processing capacity, and
communication overhead for transmitting or storing sensory samples.

To account for this limitation, we associate each slot\footnote{\label{fn:3gpp_bmg}A time slot here denotes the scheduling opportunity for
sensing and beam decision-making, distinct from the 3GPP NR
physical-layer slot~\cite{3gpp38211}, whose duration is
numerology-fixed (0.125--1~ms). NR beam management instead operates
via RRC-configured SSB/CSI-RS periodicities (5--160~ms)~\cite{3gpp38213,xue2024beammgmt}.
Our accumulated and uniform sampling policies, introduced later, mirror this periodic scheduling.}
with a binary
sensing decision $\alpha(t) \in \{0,1\}$: $\alpha(t)=1$ indicates that
sensing is performed at slot $t$ and a fresh sample is acquired for
beam prediction, while $\alpha(t)=0$ indicates that no sensing is
performed, so the predictor must reuse a previously acquired, and
hence stale, sample. This restriction on how often $\alpha(t)=1$ can
occur is formalized as the average sensing-rate constraint
later in~\eqref{eq:budget}. When more than one sensing modality is
available~\cite{Ahmed_vision}, $\alpha(t)$ is treated as a joint decision covering all
modalities.

To quantify data staleness, we employ AoI
\cite{Roy_2012, AoI_Monograph_Modiano}, defined as the number of slots elapsed since the most
recent sensing event. Given the sensing decisions, the AoI $\delta(t)$
evolves as
\begin{equation}
  \delta(t+1) = \begin{cases} 0, & \text{if } \alpha(t)=1,\\
  \delta(t)+1, & \text{if } \alpha(t)=0. \end{cases}
  \label{eq:aoi}
\end{equation}
Thus, $\delta(t)$ resets to zero whenever a fresh sample is acquired
and increases by one otherwise. When $\alpha(t)=0$, the beam predictor
at slot $t$ uses the sample acquired $\delta(t)$ slots earlier. As
illustrated in~\cite[Fig.~1]{zakeri_icc26}, the cross-entropy beam prediction loss
behaves in general as a non-monotonic function of $\delta(t)$, which motivates
modeling data freshness explicitly within the deep learning pipeline.
\subsection{Problem Formulation}
Let $g_{\boldsymbol\theta}(\cdot,\cdot)$ denote the beam predictor, parameterized by
$\boldsymbol \theta$ and detailed in \Cref{sec_method}, which maps the most recently
retrieved observation $X(t-\delta(t))$ and its age $\delta(t)$ to a
(codebook-based) beam index selection, $m(t)$, i.e., 
\begin{equation}
  m(t) = g_{\boldsymbol\theta}\big(X(t-\delta(t)),\,\delta(t)\big) \in \{1,\ldots,M\}.
  \label{eq:mt}
\end{equation}
The per-slot loss is then defined as
\begin{equation}
  f(t) := \ell\big(m(t), y(t)\big),
  \label{eq:ft}
\end{equation}
where $y(t)$ is the ground-truth optimal beam label at slot $t$,
determined offline during dataset construction, and
$\ell(\cdot,\cdot)$ is a loss function penalizing deviation of the
predicted beam from $y(t)$, e.g., cross-entropy. This loss serves as a proxy for communication
performance: it is small when the predicted beam yields a received SNR
close to that of the optimal beam, and large when the beam is
misaligned and the resulting SNR, and hence the achievable rate,
degrades. Since $m(t)$ depends on $\delta(t)$ both directly and
through the retrieved sample $X(t-\delta(t))$, $f(t)$ is a function of 
$\delta(t)$.

Over a finite horizon of $T$ time slots, the beam prediction task is
formulated as a constrained optimization problem over the joint
sequence of beam selections $\{m(t)\}_{t=1}^T$ and sensing decisions
$\{\alpha(t)\}_{t=1}^T$:
\begin{subequations}
\label{op_main}
\begin{align}
  \operatorname*{minimize}_{\{m(t),\alpha(t)\}_{t=1,2,\ldots}} \quad
    & \sum_{t=1}^T \mathbb{E}\{f(t)\} \label{eq:obj}\\
  \text{subject to} \quad
    & \frac{1}{T}\sum_{t=1}^T \mathbb{E}\{\alpha(t)\} \leq \alpha^{\max},
    \label{eq:budget}
\end{align}
\end{subequations}
where $\alpha^{\max} \in (0,1]$ bounds the fraction of slots in which
sensing may be performed. This budget represents the \textit{aggregate} and normalized cost of
sensing, which in practice may arise from power, bandwidth, or
processing limitations. The expectation is taken with respect to the possibly
randomized sensing decisions.

Problem~\eqref{op_main} couples two subproblems. The first is the \emph{sensing decision} problem: choosing \textbf{when to sense}, i.e., the sequence $\{\alpha(t)\}$, subject to the budget~\eqref{eq:budget}, referred to as ``sampling policy". The second is the \emph{beam prediction} problem: choosing 
\textbf{which beam to use} given the available data at slot $t$, i.e., $m(t)$.
 The two are coupled because the sensing decisions determine which samples the predictor observes, so future observations depend on past sensing actions. This coupling makes the joint problem a difficult sequential~decision-making~problem.

Rather than solving the joint problem above, we fix the sampling policy to
one of three budget-satisfying families (accumulated, uniform, and
randomized sampling, detailed in \Cref{sec_sampling}), each of which
satisfies constraint~\eqref{eq:budget}. 
As noted in \Cref{fn:3gpp_bmg}, accumulated and uniform sampling reflect
the periodic, centrally-scheduled cadences characteristic of
standardized beam management, while randomized sampling reflects the
irregular sensing availability induced by scheduling contention or
resource-sharing constraints. This breaks the coupling and lets us
focus on designing a beam predictor that is robust to the resulting
sparse and stale observations. This decoupling thus provides a principled
and practically relevant baseline, and we show that the choice of
sampling policy has a measurable but secondary impact compared to the
design of the predictor itself.

 \section{Age-Aware Beam Prediction Framework}\label{sec_method} 
\begin{figure}[t!]
    \centering
    \includegraphics[width=1.05\linewidth]{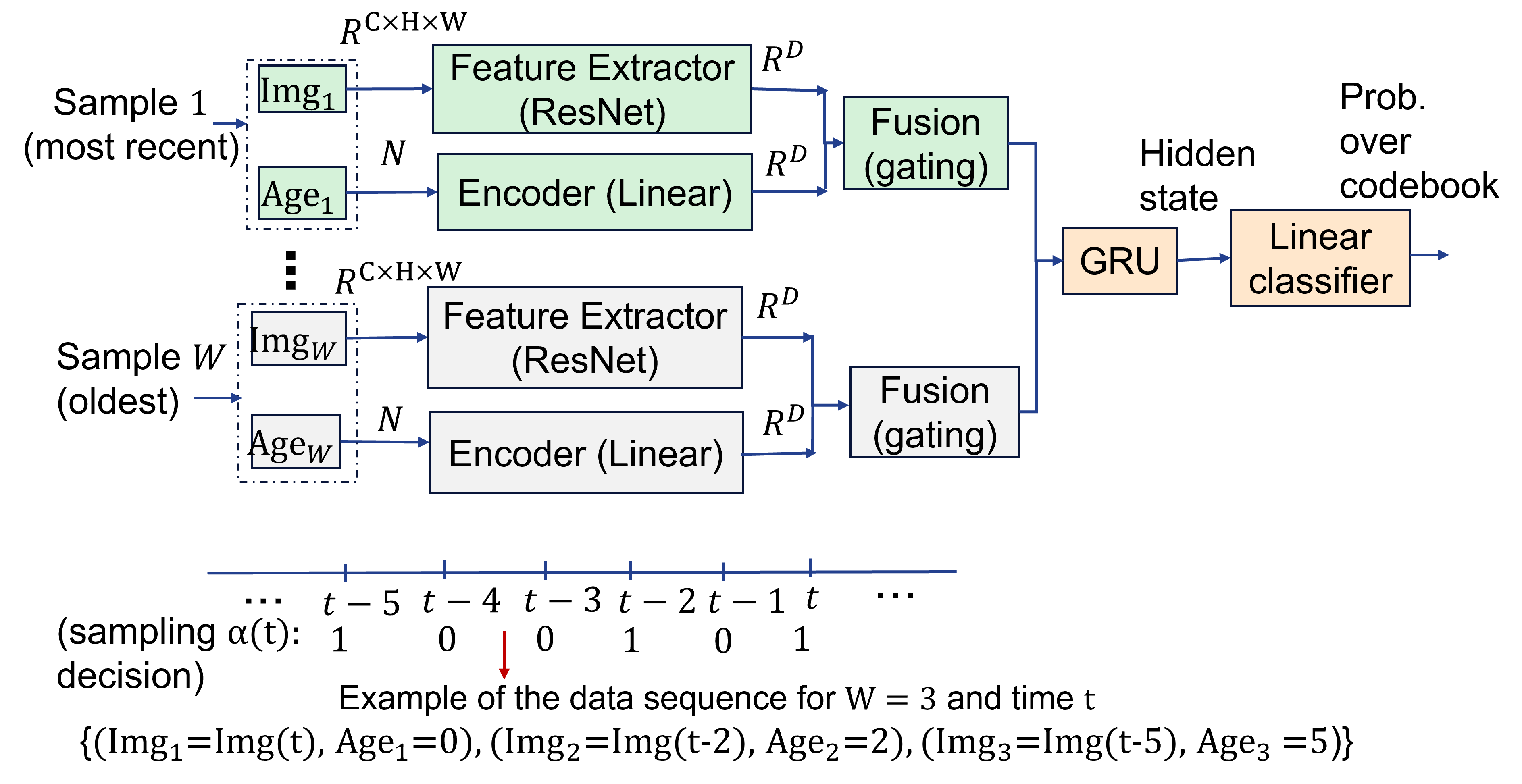}
    \caption{Proposed beam predictor architecture and data pipelines for beam prediction, where the buffered samples are ordered from most recent to oldest with non-decreasing age, and $W$ is the number of previously taken samples, i.e., the GRU sequence length.
    }
    \label{fig_NN_model}
\end{figure}

Given the sensing constraint~\eqref{eq:budget}, fresh
sensory data is not available at every time slot. The beam
predictor must therefore operate on a finite history of previously
acquired samples. In general, each sample may comprise heterogeneous
sensing modalities such as RGB images, LiDAR point clouds, or radar
measurements. In this work, we instantiate the framework with RGB
images. Let $\mathcal{T}(t)$ denote the set of the $W$ most recent
slots up to $t$ at which sensing occurred,
\begin{align}
    \mathcal{T}(t) = \big\{\, t_1 > t_2 > \cdots > t_W \;:\;
    t_w \le t,\ \alpha(t_w) = 1 \,\big\},
    \label{eq_hist_slots}
\end{align}
and let the corresponding observation history be
\begin{align}
    \mathcal{X}(t) = \big\{\, \big(\mathbf{X}(t_w),\, t - t_w\big)
    \big\}_{w=1}^{W},
    \label{eq_hist}
\end{align}
as exemplified in Fig.~\ref{fig_NN_model}. Note that the slots in
$\mathcal{T}(t)$ are in general non-consecutive, in contrast to prior
works, where the history comprises $W$ \textit{consecutive}
observations. Rather than treating data staleness as a degradation to
be tolerated, we propose to make it explicit to the beam predictor:
the age of each retrieved sample is encoded and fused with its visual
features as a synthetic modality, providing the predictor with direct
context about the reliability of its inputs. The remainder of this
section describes the complete framework, beginning with the sampling
algorithms~(Section~\ref{sec_sampling}), the preprocessing and age
encoding pipeline, and the multimodal fusion
mechanism~(Section~\ref{sec_preproc}), and the neural network
architecture~(Section~\ref{sec_nn_arch}).

\subsection{Sensing Data Sampling Algorithms}\label{sec_sampling}
As discussed in Section~\ref{sec_sm}, we restrict attention to fixed,
signal-agnostic sampling policies that satisfy the average sensing
budget constraint~\eqref{eq:budget} by construction. Here,
signal-agnostic means that the sensing decision $\alpha(t)$ does not
depend on the content of the sensed data or on the
modality~\cite{Yin_Sun_2}. This decoupling keeps the analysis
tractable and provides interpretable benchmarks against which the
benefit of the proposed age-aware learning can be isolated. We
consider three such policies.

\textit{(1) Accumulated Sampling:} We employ a deterministic
fractional-accumulator, Bresenham-style sampling scheme to ensure the
sensing budget $\alpha^{\max} \in (0, 1]$ is met exactly over any
finite horizon $T$. At each slot $t$, the accumulator state is updated and
the sensing decision is taken as
\begin{align}
    s(t) &= s(t-1) + \alpha^{\max}, \\
    \alpha(t) &= \mathds{1}\{ s(t) \ge 1 \},
\end{align}
where $\mathds{1}\{\cdot\}$ is the indicator function, and the
accumulator is reset via $s(t) \leftarrow s(t) - 1$ whenever
$\alpha(t) = 1$. With $s(0) = 0$, this construction generates the
binary sequence
\begin{align}
    \alpha(t) = \lfloor t\alpha^{\max} \rfloor -
    \lfloor (t-1)\alpha^{\max} \rfloor,
    \label{eq_bresenham}
\end{align}
which satisfies $\sum_{t=1}^{T} \alpha(t) = \lfloor T\alpha^{\max}
\rfloor$, thereby guaranteeing that the time-average sensing rate
satisfies~\eqref{eq:budget} over any horizon $T$.

\textit{(2) Uniform Sampling:} Under the average budget
$\alpha^{\max}$, sensing is performed once every $N :=
\lceil 1/\alpha^{\max} \rceil$ slots, i.e.,
\begin{align}
    \alpha(t) = \mathds{1}\{ t \bmod N = 0 \},
    \label{eq_uniform}
\end{align}
which yields a time-average sensing rate of $1/N \le \alpha^{\max}$
and hence satisfies the budget. A key structural property of this
policy is that $N = 2$ for all $\alpha^{\max} \in (0.5, 1)$, so the
selected frame set is identical to that obtained at $\alpha^{\max} =
0.5$: the fixed uniform grid cannot be refined beyond one sample every
two slots. Any budget allocated beyond this threshold is therefore
redundant, yielding no additional observations and no performance
gain. This saturation is also reflected in the numerical results of
Section~\ref{sec_nr}.

\textit{(3) Random Sampling:} We use a Bernoulli sampling scheme with
mean $\alpha^{\max}$, i.e., at every slot the sensing decision is drawn
independently as
\begin{align}
    \alpha(t) \sim \mathrm{Bernoulli}(\alpha^{\max}).
    \label{eq_bernoulli}
\end{align}
The budget constraint is thus satisfied in \textit{expectation}, with
the time-average sampling rate converging to $\alpha^{\max}$ as $T$
grows large by the law of large numbers. For short horizons, however,
individual realizations may deviate from the budget.
\subsection{Data Preprocessing, Age Encoding, and Fusion}\label{sec_preproc}
Raw RGB images are preprocessed prior to feature extraction to ensure
compatibility with pretrained convolutional backbone architectures.
Each image is resized and converted to a tensor with pixel intensities
scaled to $[0, 1]$, followed by per-channel normalization using
ImageNet statistics. This
aligns the input distribution with that encountered during backbone
pretraining and stabilizes optimization.

The AoI value $\delta(t)$ associated with each retrieved sample is
incorporated as a synthetic input signal alongside the visual data.
Unlike standard sensing modalities, $\delta(t)$ carries no
environmental information of its own; rather, it acts as a
\emph{contextual filter} that informs the predictor about the
reliability of the input image. A key practical advantage is that this
signal is available at zero sensing cost, since the AoI is computed
locally from the sampling decisions $\{\alpha(t)\}$
via~\eqref{eq:aoi}, without any additional measurement overhead.
However, as a scalar, $\delta(t)$ cannot be directly combined with
high-dimensional image feature vectors without introducing a
distributional mismatch. To address this, $\delta(t)$ is passed
through a lightweight age~encoder,
\begin{align}
    \mathbf{e}_{\mathsf{a}} =
    \sigma\big( \mathbf{w}_{\mathsf{a}} \delta(t) +
    \mathbf{b}_{\mathsf{a}} \big) \in \mathbb{R}^{D},
    \label{eq_age_enc}
\end{align}
where $\mathbf{w}_{\mathsf{a}}, \mathbf{b}_{\mathsf{a}} \in
\mathbb{R}^{D}$ are trainable parameters and $\sigma(\cdot)$ denotes
the element-wise sigmoid function. This projects the scalar age into a
$D$-dimensional embedding matching the dimension of the image feature
representation. The image embedding $\mathbf{e}_{\mathsf{i}} \in
\mathbb{R}^D$, extracted by the backbone feature extractor, and
$\mathbf{e}_{\mathsf{a}}$ are independently normalized prior to fusion
to mitigate scale~disparity.

\textit{Age Fusion:} Different multimodal fusion strategies exist,
ranging from concatenation, element-wise addition, and element-wise
multiplication to attention-based mechanisms~\cite{fusion_survey}. We
adopt early fusion, combining the two embeddings before the temporal
aggregation stage. The choice of mechanism is dictated by the
asymmetric role of the two embeddings. A symmetric strategy such as
addition or concatenation would treat the age as an equal contributor
to the fused representation, regardless of its actual relevance, which
varies substantially across sensing budgets. Element-wise
multiplication better reflects the modulating role of age, but applies
a fixed interaction that cannot adapt to how informative the age
signal is at a given budget. We therefore adopt a gating
mechanism~\cite{WenTan2025_gatefus}, in which $\mathbf{e}_{\mathsf{a}}$
modulates the image features through a learned, dimension-wise
weighting.\footnote{Although instantiated here with RGB image
features, the gating formulation is agnostic to the specific modality
and extends readily to other high-dimensional feature representations,
such as those arising from LiDAR or radar inputs.} Specifically, the
two embeddings are concatenated as $\mathbf{z} =
[\mathbf{e}_{\mathsf{i}}; \mathbf{e}_{\mathsf{a}}] \in
\mathbb{R}^{2D}$ and passed through a learnable linear transformation
followed by a sigmoid activation to produce a gating vector:
\begin{align}
    \mathbf{g} = \sigma(\mathbf{W}\mathbf{z} + \mathbf{b}),
    \label{eq:gate}
\end{align}
where $\mathbf{W} \in \mathbb{R}^{D \times 2D}$ and $\mathbf{b} \in
\mathbb{R}^D$ are trainable parameters. The fused representation is
then obtained as a dimension-wise convex combination of the two
embeddings:
\begin{align}
    \mathbf{e}_{\mathsf{fused}} = \mathbf{g} \odot
    \mathbf{e}_{\mathsf{i}} + (\mathbf{1} - \mathbf{g}) \odot
    \mathbf{e}_{\mathsf{a}},
    \label{eq:fusion}
\end{align}
where $\odot$ denotes element-wise multiplication. Since $\mathbf{g}$
in~\eqref{eq:gate} is itself a function of both embeddings, the
weighting adapts to the input rather than being fixed a priori. This
enables the model to learn how much to rely on visual content versus
age context for each feature dimension independently, providing
robustness against modality imbalance across varying sensing budgets.
A numerical comparison of gating fusion against additive,
concatenation, and multiplicative alternatives is presented in
Section~\ref{sec_nr}.

\subsection{Beam Predictor Model Architecture} \label{sec_nn_arch}
This subsection describes the neural network architecture of the
proposed beam predictor $g_{\boldsymbol{\theta}}(\cdot,\cdot)$ defined
in~\eqref{eq:mt}, detailing its constituent blocks and the two
feature extractor designs that trade off prediction accuracy against
model complexity.

As illustrated in Fig.~\ref{fig_NN_model}, the model comprises five
functional blocks: a visual feature extractor, an age encoder, a
gating fusion layer, a GRU temporal aggregator, and a beam classifier.
For each sample in the history buffer, the visual feature extractor
maps the RGB image into a feature embedding $\mathbf{e}_{\mathsf{i}}$,
while the age encoder maps the associated AoI value into an embedding
$\mathbf{e}_{\mathsf{a}}$, as described in the preceding section. The
two embeddings are combined by the gating fusion layer into a fused
representation $\mathbf{e}_{\mathsf{fused}}$. The fused embeddings
across the sampled history are then aggregated by the GRU, whose final
hidden state is mapped by the beam classifier to a probability
distribution over the $M$ codebook beams.

\textit{Visual feature extractor:} The visual feature extractor is the
most computationally intensive block of the model, as it maps each
high-dimensional RGB image into a discriminative embedding. The
remaining blocks, by contrast, operate on low-dimensional vectors.
Deep residual CNNs, and ResNet in particular, are among the most
widely adopted architectures for visual feature encoding, owing to
their strong representational capacity and the stable optimization
afforded by residual connections~\cite{resnet}. We therefore adopt a
pretrained ResNet18 backbone, with the final classification layer
removed, as the primary feature extractor. It yields a $D = 512$
dimensional embedding $\mathbf{e}_{\mathsf{i}}$ from each input image.
Although pretrained weights are used as initialization, all backbone
parameters are updated during training according to the task loss,
allowing the model to adapt its representations to the beam prediction
objective. This design contains approximately $11$ million trainable
parameters, dominated by the ResNet18 backbone.

\begin{figure}[t!]
    \centering
    \hspace{-1 em}
    \includegraphics[width=1.05\linewidth]{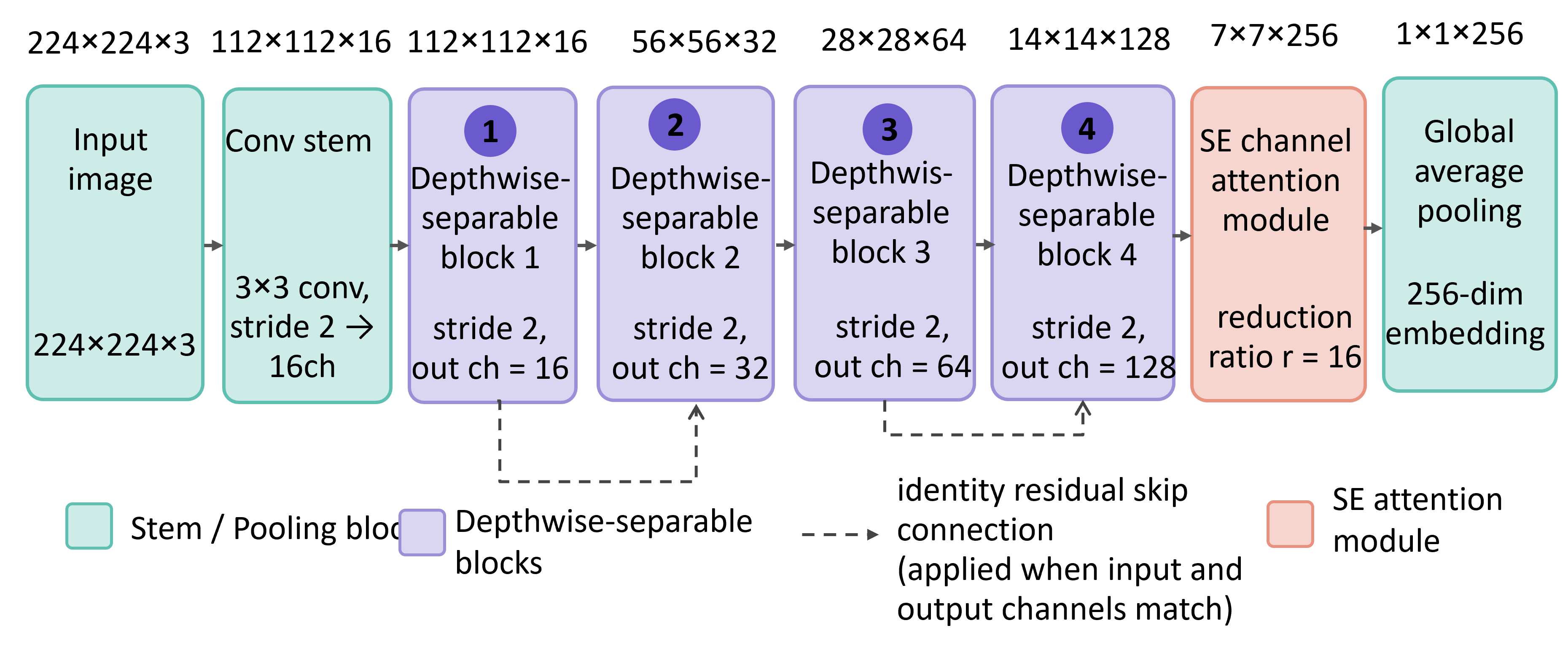}
    \caption{Our low-complexity custom CNN-based visual feature extractor model (i.e., the student model) with only around 60$K$ trainable parameters. This model replaces the heavy visual feature-extraction ResNet block in Fig.~\ref{fig_NN_model}.
    }
    \label{fig_std_model}
\end{figure}

While accurate, the ResNet18-based model is too heavy for deployment
on resource-constrained platforms, particularly since the feature
extractor accounts for nearly the entire model complexity. To address
this, we design a low-complexity visual feature extractor based on a
custom compact CNN, schematically illustrated in
Fig.~\ref{fig_std_model}. The encoder consists of a strided $3 \times
3$ convolutional stem followed by five depthwise-separable convolution
blocks with progressive channel expansion ($16 \to 32 \to 64 \to 128
\to 256$) and spatial downsampling via strided convolutions, yielding
a $256$-dimensional embedding. Identity residual connections are
applied where input and output channel dimensions match. Prior to
global average pooling, a squeeze-and-excitation (SE) channel
attention module~\cite{hu2018squeeze} with reduction ratio $r = 16$
recalibrates channel-wise feature responses. The resulting encoder
contains approximately $60$K trainable parameters, a compression ratio
of roughly $180\times$ relative to ResNet18. Since the remaining
blocks of the pipeline are inherently lightweight, replacing the
feature extractor alone substantially reduces the overall model
complexity.

Reducing model capacity in this manner, however, risks degrading
prediction accuracy, particularly under the sparse and stale inputs
induced by the sensing budget. To retain high accuracy in the compact
model, we employ a knowledge distillation (KD) framework in which the
high-capacity ResNet18-based model guides the training of the
low-complexity model. The two designs thus serve as the ``teacher''
and the ``student'', respectively; the framework is detailed in
Section~\ref{sec_kd}.

\textit{GRU temporal aggregator and classifier:} The fused embeddings
$\{\mathbf{e}_{\mathsf{fused}}\}$ of the samples in the history buffer
are fed sequentially into the GRU, ordered from oldest to most recent,
which aggregates the age-tagged observation sequence and captures
temporal dependencies across sparsely sampled frames. The final hidden
state of the GRU is passed to a fully connected classifier layer that
maps it to the logits $\{z_m(t)\}_{m=1}^{M}$ over the codebook beams,
from which the predicted distribution follows via~\eqref{eq:softmax}.
As illustrated in Fig.~\ref{fig_NN_model}, for a window size $W = 3$
at time $t$, the model processes the three most recently acquired
samples at times $t$, $t-2$, and $t-5$, with respective ages $0$, $2$,
and $5$.

A special case arises when $W = 1$, i.e., no history buffer is
maintained. In this regime, the GRU aggregation is unnecessary, and
the fused embedding $\mathbf{e}_{\mathsf{fused}}$ is passed directly to
the classifier. This reduces model complexity in two aspects: the
sequential processing of an observation history is eliminated, and the
GRU parameters are removed entirely. The empirical
implications of varying $W$ are examined in~Section~\ref{sec_nr}.

\section{Knowledge Distillation Framework}
\label{sec_kd}
\subsection{Motivation}
In the standard KD paradigm~\cite{Hinton_KD}, a compact model, called
the student, is trained to mimic the output distribution of a larger,
more capable teacher, transferring generalizable knowledge while
substantially reducing parameter count and inference cost. In most
existing applications, both teacher and student operate under
identical training and inference conditions, and the primary benefit is model compression.

Our setting differs from this paradigm due to
constraint~\eqref{eq:budget}. Specifically, the teacher is
allowed to be trained on fully sampled, temporally dense data, while
the student must train and operate under the sensing budget. Without
distillation, the student's training distribution is dominated by
high-AoI samples, especially at low budgets $\alpha^{\max}$, which prevents it from developing the inter-class structure needed to
generalize across budgets. Conversely, training a model on fully
sampled data and subsequently deploying it under a constrained
sensing budget introduces a substantial mismatch between the training
and inference distributions, leading to significant performance
degradation irrespective of model architecture or capacity. To
address both challenges, we propose to employ KD as follows: the
high-capacity teacher is first trained on \textit{fully} sampled
data, i.e., regardless of constraint~\eqref{eq:budget}, to
obtain a strong representation of the inter-class structure, which is
then transferred to the budget-constrained student through
soft-target supervision. Distillation therefore acts here as a
regularization mechanism against the distribution shift between
training and inference imposed by the sensing constraint.

To evaluate this approach, we consider two regimes for training the
teacher: (1) \emph{full sampling} (FS) and (2) \emph{limited
sampling} (LS). In the FS regime, the teacher is trained with
$\alpha(t) = 1$ for all times $t$ in the training set, i.e., each
best beam label at $t$ is paired with its associated sample
$\mathbf{X}(t)$. In the LS regime, the teacher is trained under the
sensing budget $\alpha^{\max}$, so that each best beam label at $t$
is paired with the most recent available sample
$\mathcal{X}(t-\delta(t))$, as determined by the sampling policies of
Section~\ref{sec_sampling}. Note that in both regimes the number of
slots requiring a beam decision equals the total number of sample points at inference. In both regimes, the student is trained and evaluated
exclusively under the budget~constraint.
\begin{remark}
The training regime that is optimal for a standalone predictor differs from the one that is optimal for a teacher used in distillation. When a model serves as the beam predictor itself, LS training is preferable, since training under sampling aligns the training distribution with the budget-constrained inference distribution. An FS-trained model, by contrast, overfits to dense observations and generalizes poorly to sparse inputs at inference (see results in Section~\ref{sec_nr_wkd}). When the same model instead serves as a teacher, FS training yields a stronger student (see results in Section~\ref{sec_nr_kd}), because the student already matches the inference distribution through its own budget-constrained inputs, so the teacher contributes most by supplying the rich inter-class structure learned from dense data, which the student cannot obtain on its own. This contrast confirms that, in the constrained sensing setting, KD operates as a bridge between the dense training conditions of the teacher and the sparse inference conditions of the student, rather than as a conventional compression mechanism.
\end{remark}
Having established the role of distillation, we now define the loss functions used for training. Training any model variant requires a task loss measured against the ground-truth beam labels, defined in the next subsection. Training the student additionally requires a distillation loss measured against the teacher's outputs, defined in Section~\ref{sec_kd_distloss}. For the student, the two are combined into a single KD loss.
\subsection{Task Loss}
As the beam prediction task is a multiclass classification problem,
we adopt the cross-entropy loss as the base training objective.
Let $z_m(t)$ denote the $m$-th logit of the model output at slot $t$.
Let $p_m(t)$ denote the corresponding predicted probability that
beam $m$ is optimal, obtained via the softmax function as
\begin{equation}
    p_m(t) = \mathrm{softmax}(z_m(t)) :=
    \frac{e^{z_m(t)}}{\sum_{j=1}^{M} e^{z_j(t)}},
    \quad m = 1, \ldots, M.
    \label{eq:softmax}
\end{equation}
The cross-entropy loss at slot $t$ is then
\begin{equation}
    L(t) := -\log\big(p_{y(t)}(t)\big),
    \label{eq:ce}
\end{equation}
where $y(t)$ is the ground-truth optimal beam index at $t$. 
Training is performed on minibatches of $B$ \textit {consecutive} slots. For a
minibatch $\mathcal{B}$ with $|\mathcal{B}| = B$, the empirical task
loss minimized at each training step is the sample average value given by
\begin{equation}
    \mathcal{L}^{\mathsf{task}}(\mathcal{B}) :=
    \frac{1}{B} \sum_{t \in \mathcal{B}} L(t)
    = -\frac{1}{B} \sum_{t \in \mathcal{B}}
      \log\big(p_{y(t)}(t)\big).
    \label{eq:task}
\end{equation}
This loss is used to train \textit{all} model variants, i.e., the teacher
under both the FS and LS regimes and the student without~KD.
\subsection{Distillation Loss}\label{sec_kd_distloss}
To train the student, the task loss in~\eqref{eq:task} is combined
with a distillation loss that aligns the student's output
distribution with that of the pretrained teacher. Let
$\mathcal{L}^{\mathsf{dis}}$ denote the distillation loss. The total
KD training objective over a minibatch $\mathcal{B}$ is
\begin{equation}
    \mathcal{L}^{\mathsf{kd}} :=
    \gamma \mathcal{L}^{\mathsf{dis}} +
    (1 - \gamma)\mathcal{L}^{\mathsf{task}}(\mathcal{B}),
    \label{eq:kd_loss}
\end{equation}
where $\gamma \in [0, 1]$ is a tunable hyperparameter controlling the
relative weight of the two terms.

For the distillation loss, we first consider the
Kullback-Leibler~(KL) divergence between the temperature-softened
output distributions of the teacher and student. Let
$z^{\mathsf{T}}_{m}(t)$ and $z^{\mathsf{S}}_{m}(t)$ denote the
teacher and student logits for beam $m$ at slot $t$, respectively.
The corresponding softened probabilities are ${p_m(t) =
\mathrm{softmax}\big(z^{\mathsf{T}}_{m}(t)/\tau\big)}$ and ${q_m(t) =
\mathrm{softmax}\big(z^{\mathsf{S}}_{m}(t)/\tau\big)}$, and the KL
distillation loss is given~by
\begin{equation}
    \mathcal{L}^{\mathsf{dis}}_{\mathsf{KL}} =
    \frac{\tau^2}{B} \sum_{t \in \mathcal{B}} \sum_{m=1}^{M}
    p_{m}(t) \log \left(\frac{p_{m}(t)}{q_{m}(t)}\right),
    \label{eq:kd_loss_kl}
\end{equation}
where $\tau > 1$ is the temperature parameter that softens the
distributions to expose relative class similarities, and the $\tau^2$
scaling compensates for the reduced gradient magnitude introduced by
the softening~\cite{Hinton_KD}. Substituting~\eqref{eq:kd_loss_kl}
into~\eqref{eq:kd_loss} yields the total student training objective
for this variant, which requires tuning two hyperparameters, $\tau$
and $\gamma$.

As an alternative distillation loss, inspired
by~\cite{understanding_kd, zakeri_DF_KD}, we consider the mean
squared error~(MSE) between the teacher's and student's raw logits:
\begin{equation}
    \mathcal{L}^{\mathsf{dis}}_{\mathsf{MSE}} =
    \frac{1}{B M} \sum_{t \in \mathcal{B}} \sum_{m=1}^{M}
    \left(z^{\mathsf{S}}_{m}(t) -
    z^{\mathsf{T}}_{m}(t)\right)^2,
    \label{eq:kd_loss_mse}
\end{equation}
which is substituted for $\mathcal{L}^{\mathsf{dis}}$
in~\eqref{eq:kd_loss}. Unlike the KL variant, this formulation
operates directly on the raw logits and therefore does not require
the temperature parameter $\tau$, reducing the hyperparameter search
to $\gamma$ alone. Logit matching is further motivated by its
equivalence to KL distillation in the limit of large
$\tau$~\cite{understanding_kd}, capturing inter-class structure
through the magnitude of the teacher's outputs rather than through
explicitly softened distributions. The practical implications of this
choice are examined in Section~\ref{sec_nr}.
\section{Training Algorithms}\label{sec_training}
While Section~\ref{sec_kd} defined the optimization objectives, this
section specifies the procedure by which they are optimized: we
describe the dataset construction and the application of sampling
policies across data splits, then present the training procedures for
the standard and KD-based models, summarized in
Algorithms~\ref{alg:standard_training} and~\ref{alg:kd_training},
respectively.

We adopt a time-series-based approach to dataset construction that
preserves the temporal structure of the sensing process. Given a
dataset of $N$ multimodal samples ordered chronologically, the first
$N_1$ samples are allocated for training, the following $N_2$ for
validation, and the remaining $N - N_1 - N_2$ for testing. This
chronological partitioning is essential for the sampling constraint
to be meaningful: it ensures that the history buffer
$\mathcal{X}(t)$ at any slot $t$ draws exclusively from past
observations, as it would in a real deployment, and that the test set
captures genuinely unseen future conditions. The chosen values of
$N$, $N_1$, and $N_2$ are given in~\Cref{sec_nr}.

The sampling policies introduced in Section~\ref{sec_sampling} are
applied across all three splits. For every slot $t$ at which
$\alpha(t) = 0$, the most recently acquired image sample
$\mathbf{X}(t - \delta(t))$ is retrieved and paired with its
corresponding AoI value $\delta(t)$ to form the input tuple
$(\mathbf{X}(t - \delta(t)),\, \delta(t))$. This applies uniformly to
the training, validation, and test sets, ensuring that the model is
always evaluated under conditions consistent with those it encounters
at inference. Importantly, even the teacher trained under the FS
regime is validated and tested under the sensing budget, so that
performance comparisons across all model variants remain on equal
footing.

\subsection{Training Without Distillation}
Algorithm~\ref{alg:standard_training} summarizes the training
procedure of the age-aware beam predictor model \textit{without} KD.  
The procedure is parameterized by the training mode $\mathcal{M} \in
\{\mathrm{FS},\,\mathrm{LS}\}$ defined in Section~\ref{sec_kd},
which controls whether the sensing budget is enforced during
training. In both regimes, the task loss~\eqref{eq:task} is
minimized, while validation is always performed under the sampling
budget, regardless of the training mode. Early stopping is applied
based on validation loss with patience $K$, and the model checkpoint
achieving the lowest validation loss~is~retained.

\begin{algorithm}[t!]\small
   \caption{\small Age-Aware Model Training Without Distillation}
   \label{alg:standard_training}
   \SetKwInOut{Input}{Initialize}
   \SetKwInOut{Output}{Output}
   \SetKwComment{Comment}{/* }{ */}
   \setlength{\AlCapSkip}{1em}
   \Input{
       Training set $\mathcal{D}_{\mathrm{train}}$, validation set
       $\mathcal{D}_{\mathsf{val}}$, sampling policy $\pi$, sensing
       budget $\alpha^{\max}$, training mode $\mathcal{M} \in
       \{\mathsf{FS},\,\mathrm{LS}\}$, improvement threshold $\epsilon$, max epochs $E$, patience $K$
   }

\Comment{Step (1): Construct age-tagged training sequences}
\For{each slot $t$ in $\mathcal{D}_{\mathrm{train}}$}{
    \eIf{$\mathcal{M} = \mathsf{FS}$}{
        Set $\alpha(t) \leftarrow 1$
        \tcp*{Unconstrained: always fresh data, $\delta(t)=0$}
    }{
        Draw $\alpha(t)$ from policy $\pi$ with budget $\alpha^{\max}$
        \tcp*{Budget-constrained: sparse observations}
    }
    Compute AoI $\delta(t)$ via~\eqref{eq:aoi}

    Retrieve image sample $\mathbf{X}(t - \delta(t))$

    Form input tuple $(\mathbf{X}(t - \delta(t)),\, \delta(t))$
}

\Comment{Step (2): Training loop with early stopping}
Initialize $\boldsymbol{\theta}$;\;
set $\ell^{\star} \leftarrow \infty$,\;
patience counter $c \leftarrow 0$

\For{each epoch $e = 1, \ldots, E$}{
    \For{each mini-batch $\mathcal{B}$ from
        $\mathcal{D}_{\mathrm{train}}$}{
        Compute task loss
        $\mathcal{L}^{\mathrm{task}}(\mathcal{B})$
        via~\eqref{eq:task}

        Update $\boldsymbol{\theta}$ via Adam (backpropagation)
    }
    Evaluate $\ell_{\mathsf{val}}$ on $\mathcal{D}_{\mathsf{val}}$
    under policy $\pi$ with budget $\alpha^{\max}$
    \tcp*{Inference always under sampling budget}

    \eIf{$\ell_{\mathsf{val}} < \ell^{\star} - \epsilon$}{
        $\ell^{\star} \leftarrow \ell_{\mathsf{val}}$\;
        $\boldsymbol{\theta}^{\star} \leftarrow
        \boldsymbol{\theta}$\;
        $c \leftarrow 0$
    }{
        $c \leftarrow c + 1$\;
        \lIf{$c \geq K$}{\textbf{break}}
    }
}
   \Output{Trained model parameters $\boldsymbol{\theta}^{\star}$}
\end{algorithm}

\subsection{Training With Distillation}
Algorithm~\ref{alg:kd_training} details the training procedure for
the student model under the KD framework. A key distinction from
Algorithm~\ref{alg:standard_training} is that the student is always
trained under the sensing budget, regardless of the teacher's
training regime $\mathcal{M}_{\mathsf{T}}$. At each training step,
the teacher, whose parameters are frozen, and the student process the
same age-tagged input tuple; the teacher produces soft logits that
serve as distillation targets, while the student is updated by
minimizing the combined KD loss~\eqref{eq:kd_loss}. The distillation
loss $\mathcal{L}^{\mathsf{dis}}$ is computed either as the KL
divergence~\eqref{eq:kd_loss_kl} between temperature-softened output
distributions, or as the MSE between raw
logits~\eqref{eq:kd_loss_mse}, in which case the temperature $\tau$
is not used. As with Algorithm~\ref{alg:standard_training},
validation and early stopping are always performed under the sensing
budget, ensuring that model selection reflects
inference-time~conditions.

\begin{algorithm}[t!]\small
   \caption{\small Constrained Knowledge Distillation Training (Student)}
   \label{alg:kd_training}
   \SetKwInOut{Input}{Initialize}
   \SetKwInOut{Output}{Output}
   \SetKwComment{Comment}{/* }{ */}
   \setlength{\AlCapSkip}{1em}
   \Input{
       Training set $\mathcal{D}_{\mathrm{train}}$, validation set
       $\mathcal{D}_{\mathsf{val}}$, pretrained teacher
       $f^{\mathsf{T}}(\cdot;\,\boldsymbol{\theta}^{\mathsf{T}})$
       from Alg.~\ref{alg:standard_training} with mode
       $\mathcal{M}_{\mathrm{T}} \in \{\mathsf{FS},\,\mathsf{LS}\}$,
       sampling policy $\pi$, budget $\alpha^{\max}$, KD weight
       $\gamma$, temperature $\tau$ (KL only), distillation loss type
       $\in \{\mathsf{KL},\,\mathsf{MSE}\}$, improvement threshold $\epsilon$,  max epochs $E$,
       patience~$K$
   }

\For{each slot $t$ in $\mathcal{D}_{\mathrm{train}}$}{
    Draw $\alpha(t)$ from policy $\pi$ with budget $\alpha^{\max}$
    \tcp*{Student always trains under sensing budget}

    Compute AoI $\delta(t)$ via~\eqref{eq:aoi}

    Retrieve image sample $\mathbf{X}(t - \delta(t))$

    Form input tuple $(\mathbf{X}(t - \delta(t)),\, \delta(t))$
}

\Comment{Step (2): KD training loop with early stopping}
Initialize $\boldsymbol{\theta}^{\mathsf{S}}$;\;
set $\ell^{\star} \leftarrow \infty$,\;
patience counter $c \leftarrow 0$

\For{each epoch $e = 1, \ldots, E$}{
    \For{each mini-batch $\mathcal{B}$ from
        $\mathcal{D}_{\mathrm{train}}$}{
        Get teacher logits (frozen):
        $\mathbf{z}^{\mathsf{T}}(t) \gets
        f^{\mathsf{T}}(\mathbf{X}(t{-}\delta(t)),\,\delta(t);\,
        \boldsymbol{\theta}^{\mathsf{T}})$

        Get student logits:
        $\mathbf{z}^{\mathsf{S}}(t) \gets
        f^{\mathsf{S}}(\mathbf{X}(t{-}\delta(t)),\,\delta(t);\,
        \boldsymbol{\theta}^{\mathsf{S}})$

        Compute $\mathcal{L}^{\mathsf{dis}}$ between
        $\mathbf{z}^{\mathsf{T}}$ and $\mathbf{z}^{\mathsf{S}}$
        via~\eqref{eq:kd_loss_kl} (KL)
        or~\eqref{eq:kd_loss_mse} (MSE)

        Compute total KD loss $\mathcal{L}^{\mathsf{kd}}$
        via~\eqref{eq:kd_loss}

        Update $\boldsymbol{\theta}^{\mathsf{S}}$ via Adam
        (backpropagation)
    }
    Evaluate $\ell_{\mathsf{val}}$ on $\mathcal{D}_{\mathsf{val}}$
    under policy $\pi$ with budget $\alpha^{\max}$
    \tcp*{Inference always under sampling budget}

    \eIf{$\ell_{\mathsf{val}} < \ell^{\star} - \epsilon$}{
        $\ell^{\star} \leftarrow \ell_{\mathsf{val}}$\;
        $\boldsymbol{\theta}^{\mathsf{S},\star}
        \leftarrow \boldsymbol{\theta}^{\mathsf{S}}$\;
        $c \leftarrow 0$
    }{
        $c \leftarrow c + 1$\;
        \lIf{$c \geq K$}{\textbf{break}}
    }
}
   \Output{Trained student parameters
   $\boldsymbol{\theta}^{\mathsf{S},\star}$}
\end{algorithm}

\section{Numerical Results and Discussions}\label{sec_nr}
This section presents simulation results to examine the impact of
age-augmented modality fusion, sampling algorithms, and knowledge
distillation on the beam prediction accuracy. We use the
DeepSense~6G dataset, Scenario~9, and consider a codebook of $32$
candidate beams. To account for stochastic variability, all results
are averaged over $20$ independent Monte Carlo runs, with common
random numbers (CRN)~\cite{CRN_1} used across budgets to ensure a fair comparison.

Unless stated otherwise, we set $W=1$ (single time-step input),
learning rate $\eta = 10^{-3}$, weight decay $\lambda = 10^{-4}$,
batch size $B = 64$, maximum number of training epochs $100$, a
cosine annealing scheduler with a linear warm-up of $5$ epochs (start
factor $0.1$, minimum learning rate $10^{-6}$), early-stopping
patience $K = 10$ epochs with validation loss improvement stopping threshold $\epsilon =
10^{-3}$, and the accumulated sampling policy. The overall neural
network architecture and training hyperparameters are summarized in
Table~\ref{tab_models} and Table~\ref{tab:training_hyperparams},~respectively.
\begin{table}[t]\small
\centering
\caption{Neural Network Architecture}
\label{tab_models}
\begin{tabular}{ll}
\hline
\textbf{Component} & \textbf{Value}\\
\hline
Image feature extractor   & ResNet18 (ImageNet pretrained)\\
Feature vector dimension  & 512 \\
Classifier head           & FC$(512 \rightarrow 32)$\\
Input image size          & $224 \times 224$\\
Input normalization       & Per-channel ImageNet statistics$^\dagger$\\
\hline
\multicolumn{2}{l}{$^\dagger$ Mean: $(0.485, 0.456, 0.406)$,\ Std: $(0.229, 0.224, 0.225)$.}\\ 
\end{tabular}
\end{table}
\begin{table}[t]\small
\centering
\caption{Training Hyperparameters}
\label{tab:training_hyperparams}
\begin{tabular}{ll}
\hline
\textbf{Parameter} & \textbf{Value} \\
\hline
Optimizer                      & Adam \\
Learning rate                  & $10^{-3}$ \\
Weight decay                   & $10^{-4}$ \\
Batch size                     & 64 \\
Maximum epochs                 & 100 \\
LR schedule & \makecell[l]{Linear warm-up \\ $\rightarrow$ cosine annealing} \\
Warm-up epochs                 & 5 (start factor $0.1$) \\
Minimum learning rate          & $10^{-6}$ \\
Loss function                  & Cross-entropy \\
Early-stopping patience        & 10 epochs \\
Early-stopping threshold       & $10^{-3}$ \\
Monte Carlo iterations         & 20 \\
\hline
\end{tabular}
\end{table}

\textit{Dataset Construction:} Following the chronological
partitioning described in Section~\ref{sec_training}, we allocate the
first $50$\% of samples for training, the following $15$\% for
validation, and the remaining $35$\% for testing. For Scenario~9,
with $N = 5964$ samples in total, this yields $N_1 = 3000$ training,
$N_2 = 1000$ validation, and $1964$ test samples.
\subsection{Results Without Distillation}\label{sec_nr_wkd}
Here, we first solely evaluate the impact of age fusion, sampling
algorithms, and the history window size $W$. In particular, we
consider the following benchmark and baselines.

\emph{Full sensing (upper bound)} refers to the model trained and
evaluated with $\alpha(t) = 1$ for all $t$, i.e., without the sensing
constraint~\eqref{eq:budget}, and provides a budget-independent
performance ceiling. For the proposed age-fused predictor, we
additionally compare three fusion mechanisms: additive,
concatenation-based, and multiplicative. As baselines, we
consider two without-age models that rely on the RGB image data
alone: \emph{Without age}, trained and evaluated under the sensing
budget, and \emph{Without age (FS training)}, trained with $\alpha(t)
= 1$ but evaluated under the sensing budget, which quantifies the
penalty of the training--inference distribution mismatch discussed in
Section~\ref{sec_kd}. Note that the FS regime applies only to models
without age fusion: under $\alpha(t) = 1$ the AoI is identically
zero, so the age input carries no information and the fusion
mechanism cannot meaningfully be trained. All age-fused models are therefore
trained under the sensing~budget.

\begin{figure}[t!]
\centering
\subfigure[Top-1 accuracies]
{
\includegraphics[width=0.4\textwidth]{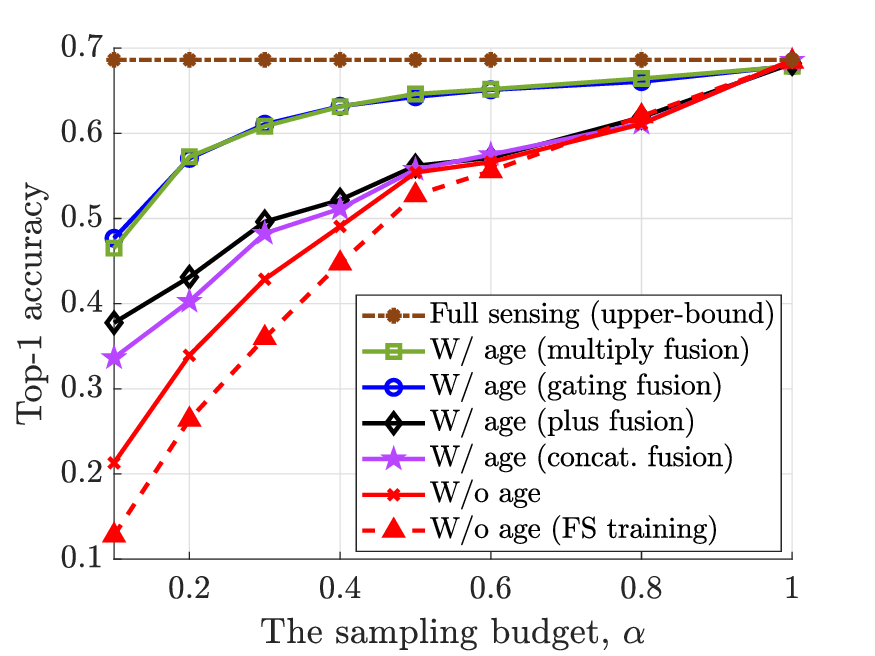}
\label{fig_fus_sc9_top1}
}

\subfigure[Top-3 accuracies]{
\includegraphics[width=0.4\textwidth]{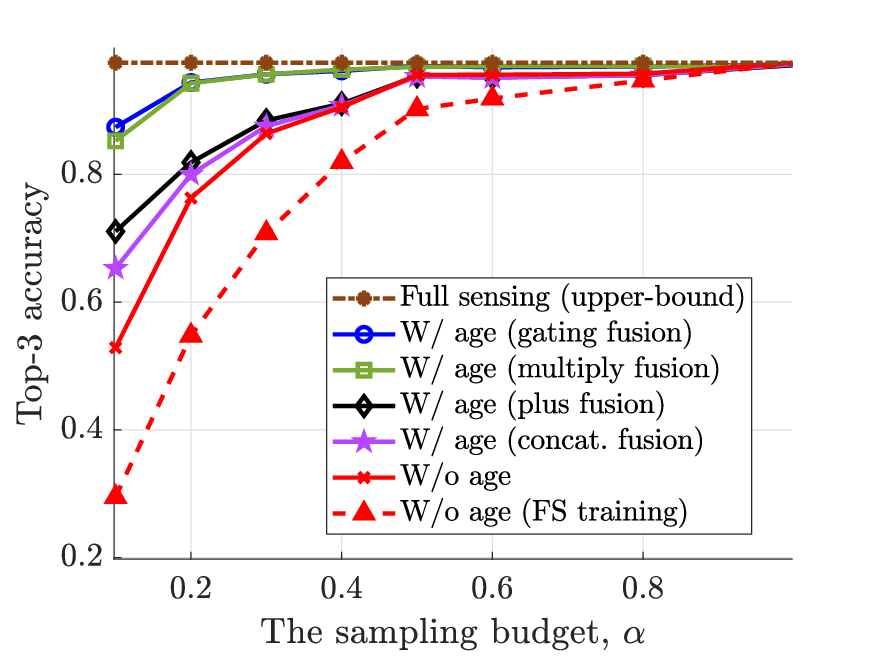}
\label{fig_fus_sc9_top3}
}
\caption{Top-1 and Top-3 accuracies for Scenario 9 across different fusion techniques. }
\label{fig_fus_sc9}
\end{figure}

\textit{Impact of Age Fusion:} Fig.~\ref{fig_fus_sc9} shows the
Top-1 and Top-3 accuracies for different fusion techniques across
sampling budgets. Age fusion yields a substantial accuracy gain: at
the most restricted budget ($\alpha^{\max} = 0.1$), the multiply and
gating fusion methods achieve a Top-1 accuracy of approximately
$47$\%, nearly doubling the $22$\% of the without age baseline.
Multiplicative and gating mechanisms further outperform additive and
concatenation-based fusion, indicating a non-linear interaction
between the age signal and the image features. This further suggests the
interpretation of age as a contextual filter on the visual features
rather than an independent data modality. While all models converge
as $\alpha^{\max} \to 1$, the age-aware models retain their advantage
in the low-budget regime, reaching near-upper-bound Top-3 accuracy
(approximately $95$\%) at $\alpha^{\max} = 0.2$, a level the baseline
does not attain until~$\alpha^{\max} = 0.5$.

\begin{figure}[t!]
 \centering
\subfigure[Top-1 accuracies]
{
\includegraphics[width=0.4\textwidth]{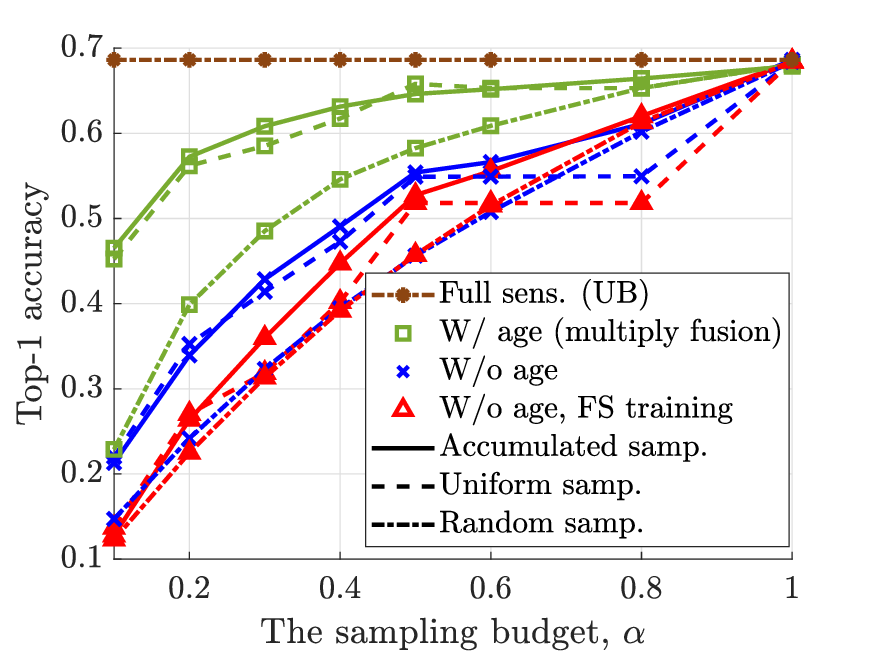}
\label{fig_sampalg_sc9_top1}
}
\subfigure[Top-3 accuracies]{
\includegraphics[width=0.4\textwidth]{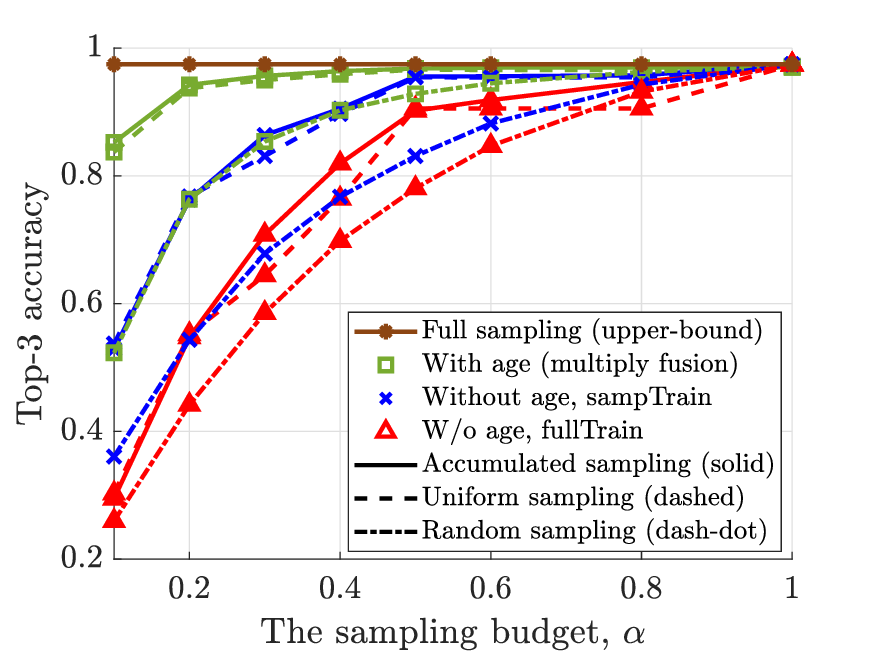}
\label{fig_sampalg_sc9_top3}
}
\caption{Top-1 and Top-3 accuracies of Scenario 9 for different sampling techniques. }
\label{fig_sampalg_sc9}
\end{figure}

\textit{Impact of Sampling Algorithms:} Fig.~\ref{fig_sampalg_sc9}
compares the three sampling policies of Section~\ref{sec_sampling},
where line style denotes the policy and marker denotes the model
variant. CRN is used to synchronize the frame subsets across model
variants at each budget, isolating the effect of the policy itself.
Accumulated sampling consistently outperforms uniform and random
sampling, with the largest gap at low $\alpha^{\max}$, as its
near-deterministic spacing yields a lower AoI variance than the
alternatives at the same average sensing rate. The three policies
converge as $\alpha^{\max} \to 1$, confirming that the gap originates
entirely from frame placement under data scarcity. A structural
limitation of uniform sampling is that for any $\alpha^{\max} > 0.5$,
the fixed uniform grid selects the same frame set as at $\alpha^{\max}
= 0.5$, so any additional budget beyond this threshold yields no
accuracy gain. The FS-trained model, in contrast, degrades sharply at
low $\alpha^{\max}$, reaching approximately $12$\% Top-1 accuracy at
$\alpha^{\max} = 0.1$: training exclusively on temporally dense
observations leaves it unable to accommodate sparse, stale inputs at
inference, consistent with the distribution mismatch discussed in
Section~\ref{sec_kd}.

The age-fused method retains its advantage over all image-only
configurations across the full budget range and under every sampling
policy, with an absolute gain of approximately $15$--$20$ percentage
points at low $\alpha^{\max}$. At low budgets, where stale
observations carry limited discriminative visual content, the age
signal provides a reliable, budget-independent cue that partially
compensates for the reduced input quality. As $\alpha^{\max}$
increases and fresher observations become available, this
contribution diminishes, since the visual features alone become
increasingly sufficient for accurate beam classification.

\begin{figure}[t!]
 \centering
\subfigure[Top-1 accuracies]
{
\includegraphics[width=0.4\textwidth]{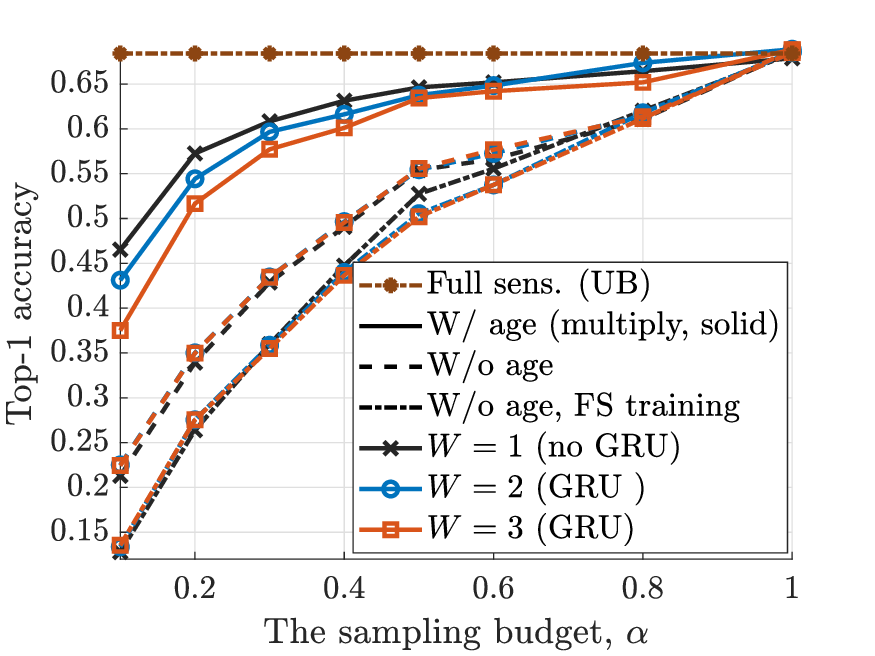}
\label{fig_gru_sc9_top1}
}

\subfigure[Top-3 accuracies]{
\includegraphics[width=0.4\textwidth]{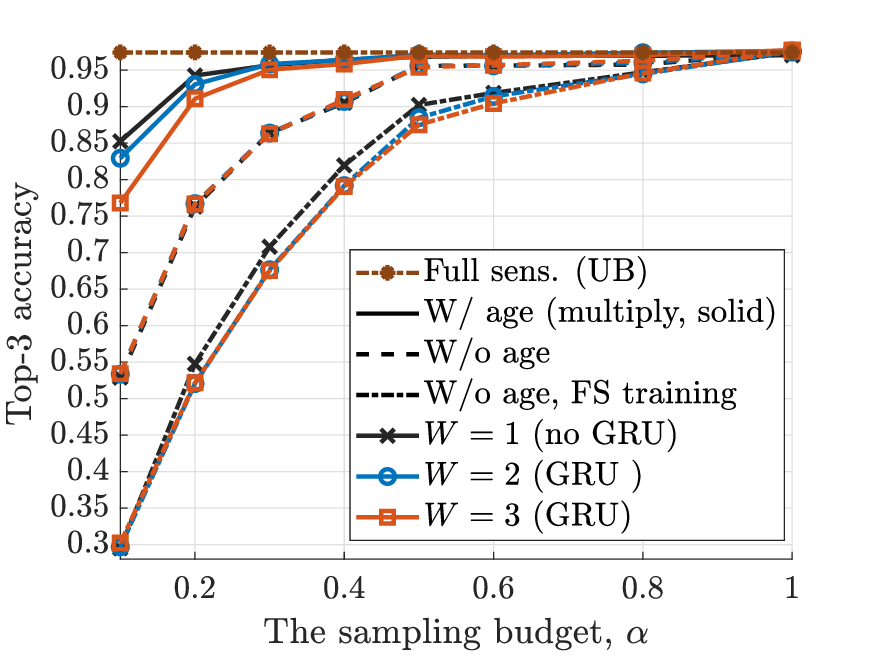}
\label{fig_gru_sc9_top3}
}
\caption{Top-1 and Top3 accuracies of Scenario 9 for the impact of the window size $W$. }
\label{fig_gru_sc9}
\end{figure}

\textit{Impact of History Window Size $W$:} Fig.~\ref{fig_gru_sc9}
examines the effect of the history window size $W$. For the age-fused
models, $W = 1$, i.e., no GRU aggregation, outperforms both $W = 2$
and $W = 3$ across the budget range, with the gap most pronounced at
low $\alpha^{\max}$ and closing only as $\alpha^{\max} \to 1$. Under
sparse sampling, the frames retained in the history window are
separated by large and variable age gaps, so aggregating over them
introduces more noise than exploitable temporal structure. This is
reinforced by the ordering of $W = 2$ and $W = 3$ at low budgets,
where the longer window performs worse, indicating that the
additional retained frames are too stale to contribute useful
context. The two converge for $\alpha^{\max} \gtrsim 0.4$, beyond
which the marginal value of an additional history step is negligible.
For the without-age models, in contrast, the ordering reverses at
mid-to-high budgets, where $W = 2$ and $W = 3$ overtake $W = 1$:
absent the age input, temporal aggregation partially substitutes for
the missing freshness information once observations are dense enough
to form a coherent sequence. The age-fused variant retains its
advantage over all image-only baselines.
\subsection{Impact of Knowledge Distillation}\label{sec_nr_kd}
Here, we examine the impact of distilling knowledge from a
high-capacity image feature encoder into the lightweight student
model described in Section~\ref{sec_nn_arch}. Unless otherwise
stated, the teacher employs the ResNet-18 backbone and the student
employs the compact custom CNN encoder, as described in
Table~\ref{tab_models}.

We consider three distilled configurations, all evaluated under the
sensing budget: \emph{FS-Img-KD} and \emph{LS-Img-KD} denote students
without age fusion, distilled from teachers trained under the FS and
LS regimes, respectively, while \emph{Age-KD} denotes the age-fused
student distilled from the FS-trained teacher. Each is compared
against the corresponding undistilled student trained directly under
the budget, with and without age~fusion.

\begin{figure}[t!]
 \centering
\subfigure[Top-1 accuracies]
{
\includegraphics[width=0.4\textwidth]{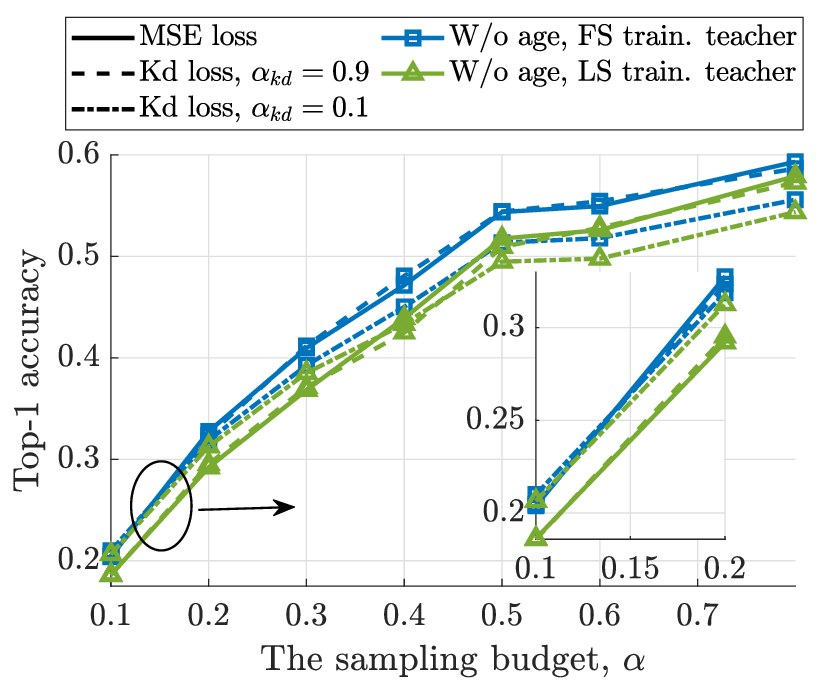}
\label{fig_kdloss_top1}
}
\subfigure[Top-3 accuracies]{
\includegraphics[width=0.4\textwidth]{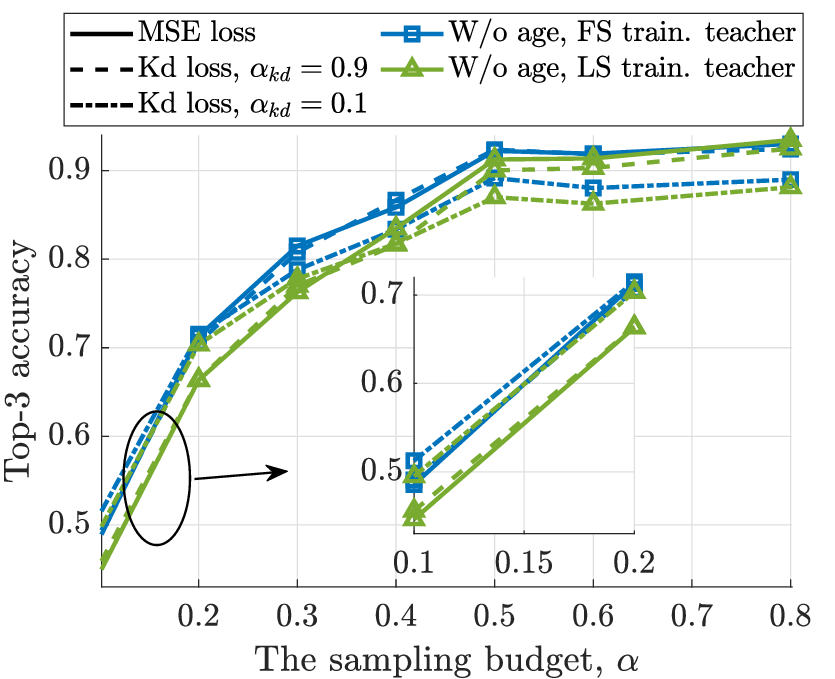}
\label{fig_kdloss_top3}
}
\caption{Top-1 and Top-3 accuracies of Scenario 9 for different KD~losses. }
\label{fig_kdloss}
\end{figure}

\textit{Effect of Distillation Loss and Teacher Training:}
Fig.~\ref{fig_kdloss} compares the MSE and KL-divergence distillation
losses (with $\gamma \in \{0.1, 0.9\}$) for the two teacher training
regimes. A cross-over is observed for the KL loss: $\gamma = 0.1$,
i.e., a higher task-loss weight, outperforms $\gamma = 0.9$ at low
budgets, while the ordering reverses at high budgets. Under severe
data scarcity, i.e., small values of $\alpha^{\max}$, anchoring the student more strongly to the
ground-truth labels is therefore beneficial, whereas at higher
budgets the richer soft-label signal from the teacher becomes more
informative and a larger distillation weight is preferable. The MSE
loss performs on par with the best KL variant across the entire
budget range while removing the temperature $\tau$ from the
hyperparameter search, making it a practical default choice for this~setting.

Fig.~\ref{fig_kdloss} further shows that the dominant factor is the
teacher's own training regime. A teacher trained under full sampling
(FS, blue curves) consistently yields a stronger student than one
trained under limited sampling (LS, green curves), across all budgets
and both Top-1 and Top-3 metrics, with the gap widening at
low~$\alpha^{\max}$ as seen in the inset. This ordering is the
reverse of the standalone comparison in Sec.~\ref{sec_nr_wkd}, where
the LS-trained teacher achieves higher accuracy under constrained
inference. The reversal confirms the mechanism described in
Section~\ref{sec_kd}: what transfers to the student is the
inter-class structure the teacher acquires from fully observed data,
not its accuracy under the budget.

\subsubsection*{Overall Impact of KD and Age Fusion}
\begin{figure}[h!]
\centering
\subfigure[Top-1 accuracies]
{
\includegraphics[width=0.42\textwidth]{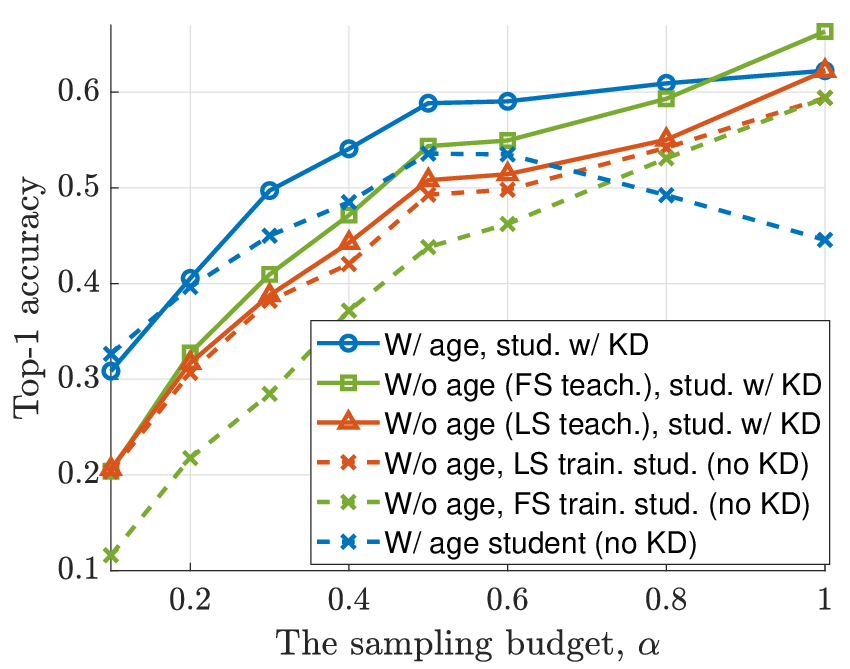}
\label{fig_kd_top1}
}
\subfigure[Top-3 accuracies]{
\includegraphics[width=0.42\textwidth]{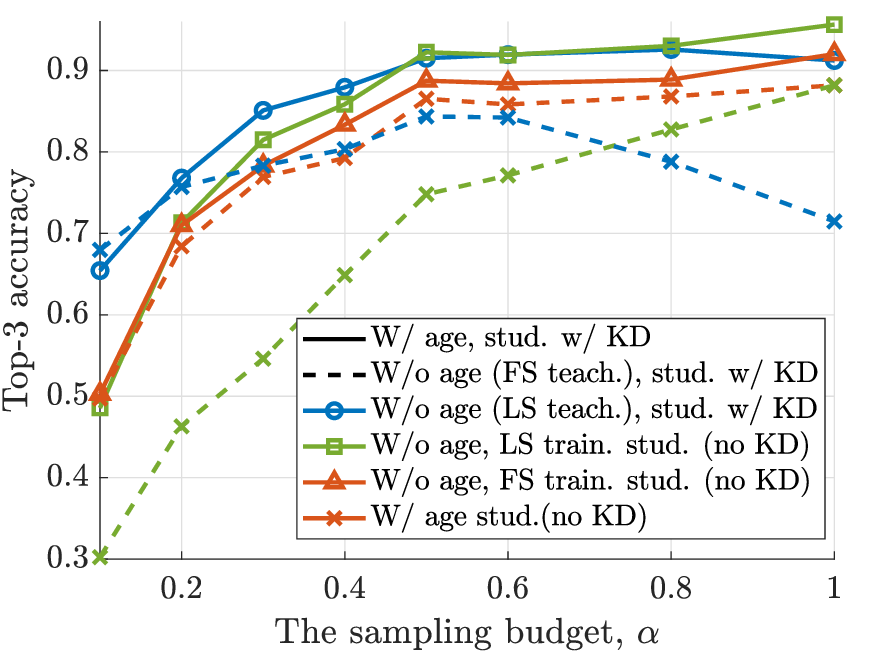}
\label{fig_kd_top3}
}
\caption{Top-1 and Top-3 accuracies of Scenario 9 for the impact~of~KD.
}
\label{fig_kd}
\end{figure}

Fig.~\ref{fig_kd} compares distilled and non-distilled student
models, with and without age fusion. KD consistently improves
accuracy over all no-KD baselines across the full budget range,
confirming that the teacher's knowledge transfers effectively to the
compact student even under sampling constraints. Among the distilled
configurations, \textit{Age-KD} achieves the highest Top-1 accuracy
throughout, with FS-Img-KD trailing marginally and both substantially
outperforming LS-Img-KD, consistent with the finding above that the
teacher's training regime is the primary~differentiator.

The \textit{W/ age student (no KD)} baseline exhibits a distinct
failure mode. At low budgets ($\alpha^{\max} \leq 0.3$) it is
competitive with Age-KD, but its accuracy stalls at mid-range budgets
and plateaus near $0.47$ Top-1 at $\alpha^{\max} = 1$, falling below
all distilled models. This indicates an optimization conflict:
without the teacher's soft-label guidance, the fusion mechanism fails
to reweight the age and image contributions as more visual data
becomes available, so the age signal interferes rather than
complements. KD resolves this by regularizing the student's output
distribution, enabling it to exploit both inputs across all budgets.

Finally, the \textit{W/o age, FS train.\ stud.\ (no KD)} variant
performs poorly at low~$\alpha^{\max}$ ($\approx 0.12$ Top-1 at
$\alpha^{\max} = 0.1$), well below all other configurations. As
discussed previously, training exclusively on fully sampled data
causes the model to overfit to dense observations and lose robustness
to sparse inference inputs. The distilled variants avoid this
collapse: the teacher's soft targets encode abstract beam-class
relationships that transfer more gracefully to low-budget conditions,
confirming the role of KD in this setting as an implicit regularizer
against the training--inference distribution shift rather than as a
compression mechanism.

\section{Conclusions}\label{sec_con}
This paper studied sensing-aided beam prediction under an explicit
average sensing rate constraint, which limits how often a data sample
can be acquired in association with a beam decision slot. The age of
the input sample is encoded and fused with the visual features
through a gating mechanism.
Training at a high sensing rate yields better beam prediction
accuracy, but such a model fails when inference is performed under
the sensing constraint. We proposed KD to address this: a teacher is
trained on fully sampled data, without accounting for the sensing
constraint, and its acquired knowledge is then distilled into a
student deployed under that constraint. This yields not only a
low-cost model for beam prediction, but a more accurate one under
sparse data availability.
Future work may consider modality-specific cost budgets in place of
the shared $\alpha(t)$; joint optimization of the sensing and prediction
policies; and extension of age-aware KD to distribution shifts beyond data sparsity at
inference, e.g., sensor noise or unseen deployment scenarios.

\bibliographystyle{ieeetr}
\bibliography{Bib_References/conf_short,
Bib_References/IEEEabrv,
Bib_References/Bibliography, Bib_References/multimodalsensing_Bio, Bib_References/ML_Bio}

@INPROCEEDINGS{Ahmed_vision,
  author={Charan, Gouranga and Osman, Tawfik and Hredzak, Andrew and Thawdar, Ngwe and Alkhateeb, Ahmed},
  booktitle=wcnc, 
  title={Vision-Position Multi-Modal Beam Prediction Using Real Millimeter Wave Datasets}, 
  year={Austin, TX, USA, Apr. 2022},
  volume={},
  number={},
  pages={2727-2731},
  doi={10.1109/WCNC51071.2022.9771835}}

@article{AoI_Monograph_Modiano,
  title={Age of information: {A} new metric for information freshness},
  author={Sun, Yin and Kadota, Igor and Talak, Rajat and Modiano, Eytan},
  journal={Synthesis Lectures on Communication Networks},
  volume={12},
  number={2},
  pages={1--224},
  year={Dec. 2019},
  publisher={Morgan \& Claypool Publishers}
}

@ARTICLE{Yin_Sun_2,  author={Bedewy, Ahmed M. and Sun, Yin and Kompella, Sastry and Shroff, Ness B.}, 
journal=IEEE_J_IT,   title={Optimal Sampling and Scheduling for Timely Status Updates in Multi-Source Networks},   year={Jun. 2021},  volume={67},  number={6},  pages={4019-4034},  doi={10.1109/TIT.2021.3060387}}

@INPROCEEDINGS{Roy_2012,  author={S. {Kaul} and R. {Yates} and M. {Gruteser}},  booktitle={Proc. IEEE Int. Conf. on Computer Commun. 
},   title={Real-time status: How often should one update?},  
year={Orlando, FL, USA, Mar. 2012},  volume={},  
number={},  pages={2731-2735},}

@STRING{IEEE_J_VT         = "{IEEE} Trans. Veh. Technol."}

@STRING{IEEE_J_JSAC       = "{IEEE} J. Sel. Areas Commun."}

@STRING{IEEE_J_WCOM       = "{IEEE} Trans. Wireless Commun."}

@STRING{IEEE_J_IT         = "{IEEE} Trans. Inf. Theory"}

@STRING{IEEE_J_IOT        = "{IEEE} Internet Things J."}

@STRING{IEEE_M_COM        = "{IEEE} Commun. Mag."}

@STRING{IEEE_O_CSTO       = "{IEEE} Commun. Surveys Tuts."}

@STRING{IEEE_M_WC         = "{IEEE} Wireless Commun."}

@inproceedings{resnet,
  author    = {He, Kaiming and Zhang, Xiangyu and Ren, Shaoqing and Sun, Jian},
  title     = {Deep Residual Learning for Image Recognition},
  booktitle = {Proc. IEEE Conf. Comput. Vis. Pattern Recognit. (CVPR)},
  address   = {Las Vegas, NV, USA},
  month     = Jun,
  year      = {2016},
  pages     = {770--778},
}

@InProceedings{hu2018squeeze,
  author    = {Hu, Jie and Shen, Li and Sun, Gang},
  title     = {Squeeze-and-Excitation networks},
  booktitle = {Proc. IEEE Conf. Comput. Vis. Pattern Recognit.},
  pages     = {7132--7141},
  month     = {June},
  year      = {2018},
  address   = {Salt Lake City, UT, USA}
}

@article{fusion_survey,
  author  = {Li, Songtao and Tang, Hao},
  title   = {Multimodal Alignment and Fusion: A Survey},
  journal = {arXiv preprint arXiv:2411.17040},
  year    = {Oct. 2025},
  url     = {https://arxiv.org/abs/2411.17040},
  howpublished = {\url{https://arxiv.org/pdf/2411.17040}}
}

@article{WenTan2025_gatefus,
  author  = {Wen, Bin and Tan, Tien-Ping},
  title   = {{PGF-Net}: A Progressive Gated-Fusion Framework for Efficient Multimodal Sentiment Analysis},
  journal = {arXiv preprint arXiv:2508.15852},
  year    = {2025},
  doi     = {10.48550/arXiv.2508.15852},
  url     = {https://arxiv.org/abs/2508.15852}
}

@article{understanding_kd,
  author    = {Taehyeon Kim and Jaehoon Oh and Nakyil Kim and Sangwook Cho and Se‑Young Yun},
  title     = {Understanding Knowledge Distillation},
  journal   = {OpenReview (ICLR Submission)},
  year      = {2019},
  url       = {https://openreview.net/forum?id=tcjMxpMJc95}
}

@article{Hinton_KD,
  title={Distilling the Knowledge in a Neural Network},
  author={Hinton, Geoffrey and Vinyals, Oriol and Dean, Jeff},
  journal={arXiv preprint arXiv:1503.02531},
  year={2015},
  url={https://arxiv.org/abs/1503.02531}
}

@string{ icc = {Proc. IEEE Int. Conf. Commun.}}

@string{ icassp = {Proc. IEEE Int. Conf. Acoust., Speech, Signal Processing}}

@string{ milcom = {Proc. IEEE Military Commun. Conf.}}

@string{ vtc = {Proc. IEEE Veh. Technol. Conf.}}

@string{ wcnc = {Proc. IEEE Wireless Commun. and Networking Conf.}}

@string{ mass_smart = {Proc. Int. Conf. Mobile, Ad Hoc, and Smart Sys. (MASS)}}

@techreport{3gpp38213,
  title       = {{NR}; Physical Layer Procedures for Control},
  institution = {3GPP},
  number      = {TS 38.213},
  year        = {2024},
  note        = {Release 18}
}

@techreport{3gpp38211,
  title       = {{NR}; Physical Channels and Modulation},
  institution = {3GPP},
  number      = {TS 38.211},
  year        = {2024},
  note        = {Release 18}
}

@article{xue2024beammgmt,
  author  = {Xue, Qing and Ji, Chengwang and Ma, Shaodan and Guo, Jiajia and Xu, Yongjun and Chen, Qianbin and Zhang, Wei},
  title   = {A Survey of Beam Management for {mmWave} and {THz} Communications Towards {6G}},
  journal = IEEE_O_CSTO,
  volume  = {26},
  number  = {3},
  pages   = {1520--1559},
  year    = {2024},
  doi     = {10.1109/COMST.2024.3361991}
}

@ARTICLE{Tan_V2X_25,
   author  = {Tan, Kang and Zhu, Ce},
   journal = IEEE_J_IOT,
   title   = {Multimodal Sensing for Intelligent {V2X}: A Review of Recent Advances Toward Deployment},
   year    = {Nov. 2025},
   volume  = {12},
   number  = {22},
   pages   = {46294--46315},
   doi     = {10.1109/JIOT.2025.3601236}
 }

@ARTICLE{Xie_contrast_bt,
   author  = {Xie, Xin and Liao, Feiyue and Wang, Heng},
   journal = {Under Review},
   title   = {A Beam Tracking Approach for {mmWave} Communication with Multi-Modal Data Fusion},
   year    = {2025}
 }

@ARTICLE{Zheng_JEPA_MSAC,
   author  = {Zheng, Can and He, Jiguang and Cai, Guofa and Li, Nannan and Bennis, Mehdi and Wymeersch, Henk and Debbah, M{\'{e}}rouane},
   journal = {Under Review},
   title   = {{JEPA-MSAC}: A Joint-Embedding Predictive Architecture for Multimodal Sensing-Assisted Communications},
   year    = {2025}
 }

@INPROCEEDINGS{Li_OFDM_mm,
   author    = {Li, Yinghan and Yu, Wei},
   booktitle = {Proc. IEEE Int. Conf. Acoust., Speech Signal Process. (ICASSP)},
   title     = {Multimodal Sensing-Aided Beamforming Optimization for {OFDM} Systems},
   year      = {2025}
 }

@ARTICLE{Cai_mapISAC,
   author  = {Cai, Xiao and Pan, Guangjin and Chen, Hui and Wymeersch, Henk and Cheng, Hei Victor},
   journal = {Under Review},
   title   = {Map-Aided {ISAC} Beam Design via Active Sensing},
   year    = {2025}
 }

@article{CRN_1,
  author  = {Kleijnen, Jack P. C. and Ridder, Ad and Rubinstein, Reuven},
  title   = {Variance Reduction Techniques in {Monte Carlo} Methods},
  journal = {SSRN Electronic Journal},
  year    = {2010},
  doi     = {10.2139/ssrn.1715474}
}

@ARTICLE{Mollah_multimodal_attention,
  author={Mollah, Muhammad Baqer and Wang, Honggang and Karim, Mohammad Ataul and Fang, Hua},
  journal=IEEE_J_VT, 
  title={Multi-Modal Sensing and Fusion in {Mmwave} Beamforming for Connected Vehicles: A Transformer Based Framework}, 
  year={Early Access, 2026},
  volume={},
  number={},
  pages={1-13},
  doi={10.1109/TVT.2026.3665294}}

@ARTICLE{ahmed_vis_tvt,
  author={Charan, Gouranga and Alrabeiah, Muhammad and Alkhateeb, Ahmed},
  journal=IEEE_J_VT, 
  title={Vision-Aided {6G} Wireless Communications: Blockage Prediction and Proactive Handoff}, 
  year={Oct. 2021},
  volume={70},
  number={10},
  pages={10193-10208},
  doi={10.1109/TVT.2021.3104219}}

@INPROCEEDINGS{zakeri_drl_sen,
  author={Zakeri, Abolfazl and Nguyen, Nhan Thanh and Alkhateeb, Ahmed and Juntti, Markku},
  booktitle=icassp, 
  title={Deep Reinforcement Learning for Dynamic Sensing and Communications}, 
  year={Barcelona, Spain, May 2026},
  volume={},
  number={},
  pages={20636-20640},
  doi={10.1109/ICASSP55912.2026.11464189}}

@article{zakeri_icc26,
  title={{AoI}-Aware Machine Learning for Constrained Multimodal Sensing and Communications},
  author={A. Zakeri and Nguyen, Nhan Thanh and Alkhateeb, Ahmed and Juntti, Markku},
  journal=icc,
  year={Accepted, May 2026}
}

@article{zakeri_DF_KD,
  title={Data-Free Knowledge Distillation for {LiDAR}-Aided Beam Tracking in {MmWave} Systems},
  author={Zakeri, Abolfazl and Nguyen, Nhan Thanh and Alkhateeb, Ahmed and Juntti, Markku},
  journal={arXiv preprint arXiv:2509.19092},
  year={Sep. 2025}
}

@article{Ma_KD,
  title={Knowledge Distillation for Sensing-Assisted Long-Term Beam Tracking in {mmWave} Communications},
  author={Ma, Mengyuan and Nguyen, Nhan Thanh and Shlezinger, Nir and Eldar, Yonina C and Swindlehurst, A Lee and Juntti, Markku},
  journal={arXiv preprint arXiv:2509.11419},
  year={Sep. 2025}
}

@ARTICLE{multimodal_wc_mag,
  author={Kim, Seungnyun and Moon, Jihoon and Kim, Jinhong and Ahn, Yongjun and Kim, Donghoon and Kim, Sunwoo and Shim, Kyuhong and Shim, Byonghyo},
  journal=IEEE_M_WC, 
  title={Role of Sensing and Computer Vision in {6G} Wireless Communications}, 
  year={Oct. 2024},
  volume={31},
  number={5},
  pages={264-271},
  doi={10.1109/MWC.016.2300526}}

@INPROCEEDINGS{multimo_revis,
  author={Vuckovic, Katarina and Hosseini, Saba M. and Rahnavard, Nazanin},
  booktitle=milcom, 
  title={Revisiting Performance Metrics for Multimodal {mmWave} Beam Prediction Using Deep Learning}, 
  year={Washington, DC, USA, Oct, 2024},
  volume={},
  number={},
  pages={881-887},
  doi={10.1109/MILCOM61039.2024.10773747}}

@INPROCEEDINGS{digital_twin,
  author={Arnold, Maximilian and Major, Bence and Massoli, Fabio Valerio and Soriaga, Joseph B. and Behboodi, Arash},
  booktitle=ICC, 
  title={Vision-Assisted Digital Twin Creation for {mmWave} Beam Management}, 
  year={Denver, CO, USA, Jun. 2024},
  volume={},
  number={},
  pages={1-6},
  doi={10.1109/ICC51166.2024.10622161}}

@INPROCEEDINGS{multmodal_exp_vtc,
  author={Li, Kehui and Zhou, Binggui and Guo, Jiajia and Yang, Xi and Xue, Qing and Gao, Feifei and Ma, Shaodan},
  booktitle=vtc, 
  title={Vision-aided Multi-user Beam Tracking for {mmWave} Massive {MIMO} System: Prototyping and Experimental Results}, 
  year={Singapore, Jun. 2024},
  volume={},
  number={},
  pages={1-6},
  doi={10.1109/VTC2024-Spring62846.2024.10683659}}

@ARTICLE{vision_aid_pos_JSAC_24,
  author={Kim, Seungnyun and Moon, Jihoon and Wu, Jiao and Shim, Byonghyo and Win, Moe Z.},
  journal=IEEE_J_JSAC, 
  title={Vision-Aided Positioning and Beam Focusing for {6G} Terahertz Communications}, 
  year={Sep. 2024},
  volume={42},
  number={9},
  pages={2503-2519},
  doi={10.1109/JSAC.2024.3413949}}

@ARTICLE{Gerhad_bt,
  author={Qurratulain Khan, M. and Gaber, Abdo and Schulz, Philipp and Fettweis, Gerhard},
  journal={IEEE Access}, 
  title={Machine Learning for Millimeter Wave and Terahertz Beam Management: A Survey and Open Challenges}, 
  year={2023},
  volume={11},
  number={},
  pages={11880-11902},
  doi={10.1109/ACCESS.2023.3242582}}

@INPROCEEDINGS{Sun_multimodal_RI,
  author={Zhang, Keyuan and Sun, Yin and Ji, Bo},
  booktitle=mass_smart, 
  title={Multimodal Remote Inference}, 
  year={Chicago, IL, USA, Oct. 2025},
  volume={},
  number={},
  pages={198-204},
  doi={10.1109/MASS66014.2025.00039}}

@ARTICLE{R_Heath_mag,
  author={Ali, Anum and Gonzalez-Prelcic, Nuria and Heath, Robert W. and Ghosh, Amitava},
  journal=IEEE_M_COM, 
  title={Leveraging Sensing at the Infrastructure for {mmWave} Communication}, 
  year={Jul. 2020},
  volume={58},
  number={7},
  pages={84-89},
  doi={10.1109/MCOM.001.1900700}}

@ARTICLE{Ahmed_deepsense,
  author={Alkhateeb, Ahmed and Charan, Gouranga and Osman, Tawfik and Hredzak, Andrew and Morais, Joao and Demirhan, Umut and Srinivas, Nikhil},
  journal=IEEE_M_COM, 
  title={DeepSense {6G}: A Large-Scale Real-World Multi-Modal Sensing and Communication Dataset}, 
  year={Sep. 2023},
  volume={61},
  number={9},
  pages={122-128},
  doi={10.1109/MCOM.006.2200730}}

@ARTICLE{RHeath_BT_multiuser,
  author={Patel, Kartik and Heath, Robert W.},
  journal= IEEE_J_WCOM, 
  title={Harnessing Multimodal Sensing for Multi-User Beamforming in {mmWave} Systems}, 
  year={Dec. 2024},
  volume={23},
  number={12},
  pages={18725-18739},
  doi={10.1109/TWC.2024.3475950}}

@article{Walid_ML25,
  title={Resource-Efficient Beam Prediction in {mmWave} Communications with Multimodal Realistic Simulation Framework},
  author={Park, Yu Min and Tun, Yan Kyaw and Saad, Walid and Hong, Choong Seon},
  journal={arXiv preprint arXiv:2504.05187},
  year={Apr. 2025}
}

\end{document}